\pdfoutput=1
\documentclass[acmsmall]{acmart}
\setcopyright{acmcopyright}
\renewcommand\footnotetextcopyrightpermission[1]{} 

\usepackage{amsthm}
\theoremstyle{remark}

\newtheorem{assumption}{Assumption}

\usepackage{paper}
\usepackage{graphicx}
\usepackage{multicol}
\usepackage{multirow}
\usepackage{bbding}
\usepackage{colortbl}
\usepackage{enumitem}
\usepackage{epstopdf}
\usepackage{bm,upgreek}
\usepackage[font=small,labelfont=bf]{caption}
\usepackage{algorithmicx}
\usepackage{algpseudocode}
\algdef{SE}[DOWHILE]{Do}{doWhile}{\algorithmicdo}[1]{\algorithmicwhile\ #1}

\usepackage{color, soul}

\definecolor{light-gray}{gray}{0.80}
\definecolor{lightblue}{rgb}{.90,.95,1}
\sethlcolor{light-gray}

\usepackage{pifont}
\newcommand{\cmark}{\ding{51}}
\newcommand{\xmark}{\ding{55}}
\newcommand{\M}{\phantom{0}}

\usepackage{fontawesome}

\newcommand\numberthis{\addtocounter{equation}{1}\tag{\theequation}}

\newcommand{\ceil}[1]{\left\lceil #1 \right\rceil}

\title{Verifix: Verified Repair of Programming~Assignments}
\begin{document}

\author{Umair Z. Ahmed}
\email{umair@nus.edu.sg}
\affiliation{%
  \institution{National University of Singapore}
  \country{Singapore}
}

\author{Zhiyu Fan}
\affiliation{%
  \institution{National University of Singapore}
  \country{Singapore}
}
\email{zhiyufan@comp.nus.edu.sg}

\author{Jooyong Yi}\thanks{Corresponding Author: Jooyong Yi}
\affiliation{%
  \institution{Ulsan National Institute of Science and
           Technology}
  \country{Korea}
}
\email{jooyong@unist.ac.kr}

\author{Omar I. Al-Bataineh}
\affiliation{%
  \institution{National University of Singapore}
  \country{Singapore}
}
\email{omerdep@yahoo.com}

\author{Abhik Roychoudhury}
\affiliation{%
 \institution{National University of Singapore}
 \country{Singapore}
}
\email{abhik@comp.nus.edu.sg}

\clubpenalty=10000
\widowpenalty = 10000
\displaywidowpenalty = 10000

\newcommand{\APRS}{APR\xspace} 

\begin{abstract}
Automated feedback generation for introductory programming assignments is useful for programming education. Most works try to generate feedback to correct a student program by comparing its behavior with an instructor's reference program on selected tests.
In this work, our aim is to generate verifiably correct program repairs as student feedback. The student assignment is aligned and composed with a reference solution in terms of control flow, and differences in data variables are automatically summarized via predicates to relate the variable names. Failed verification attempts for the equivalence of the two programs are exploited to obtain a collection of maxSMT queries, whose solutions point to repairs of the student assignment. We have conducted experiments on  student assignments curated from a widely deployed intelligent tutoring system. Our results indicate that we can generate verified feedback in up to 58\% of the assignments. More importantly, our system indicates when it is able to generate a verified feedback, which is then usable by novice students with high confidence. 
\end{abstract}

\maketitle

\lstset{frame=lines,
        frame=none,
        escapechar=@
        mathescape=false,
        language = C,
        basicstyle = \sffamily\small,
        commentstyle = \upshape\sffamily\small,
        keywordstyle = \sffamily\small,
        boxpos = t,
        xleftmargin=15pt,
        numbers=left,
        basewidth=0.5em
}

\section{Introduction}
\label{sec:intro}

\begin{table}[t]
  \centering \small
  \begin{tabular}{c|c c c}
      Tool & Completely & Beyond Identical & Verified \\
           & Automated & Reference CFG  & Repair    \\
      \hline
      Clara~\cite{clara} & \cmark & \xmark  & \xmark \\
      SarfGen~\cite{sarfgen} & \cmark & \xmark & \xmark \\
      
      ITSP~\cite{yi2017feasibility} & \cmark & \cmark & \xmark \\
      Refactory~\cite{refactory} & \cmark & \cmark & \xmark \\
      
      CoderAssist~\cite{gulwani-fse16} & \xmark & \cmark & \cmark \\
      
      \hline
      Verifix & \cmark & \cmark & \cmark \\
  \end{tabular}
  \caption{Programming assignment repair tools comparison.}
  \label{tab:tools}
\end{table}

\begin{figure*}[t] 
  \begin{subfigure}[b]{0.31\textwidth}
  
    \begin{lstlisting}[escapechar=?]
  int check_prime(int n)
  {
    if (n == 1) 
      return 0;
    int j; 
    for(j=2; j<n; j++)
    {
      if (n%j == 0)
        return 0;
      
      
    }
    return 1;
  }
    \end{lstlisting}
    \caption{The correct reference program}
    \label{fig:example-ref}
  \end{subfigure}
  \begin{subfigure}[b]{0.32\textwidth}
    \begin{lstlisting}[escapechar=@]
  int check_prime(int n)
  {
  
  
    int i=1;
    while(i<=n-1)
    {
      if (n%i != 0) 
        ; @// No-Op@
      else 
        return 0;
    }
    return 1;
  }    
     \end{lstlisting}
     \caption{The incorrect student program}
      \label{fig:example-incorrect}
  \end{subfigure}
  \begin{subfigure}[b]{0.33\textwidth}
    \begin{lstlisting}[escapechar=@]
  int check_prime(int n)
  {
    @\hl{if (n == 1)}@
      @\hl{return 0;}@
    int @\hl{i=2}@; 
    while(i<=n-1)
    {
      if (n%i != 0) 
        @\hl{i++;}@
      else 
        return 0;
    }
    return 1;
  }    
    \end{lstlisting}
    \caption{The repaired program by Verifix}
    \label{fig:example-fixed}
  \end{subfigure}
  %
    \caption{Motivating example for the \textit{Prime Number} programming assignment. Existing tools such as Clara~\cite{clara} and Sarfgen~\cite{sarfgen} cannot repair the incorrect student program in Fig~\ref{fig:example-incorrect} since its Control-Flow Graph (CFG) differs from the CFG of instructor designed reference program in Fig~\ref{fig:example-ref}. Our tool Verifix generates the repaired program in Fig~\ref{fig:example-fixed}, which is verifiably equivalent to the reference implementation, due to superior Control-Flow Automata (CFA) based abstraction.}
    \label{fig:example}
  \end{figure*}

CS-1, the introductory programming course, is an undergraduate course offered by Universities and Massive Open Online Courses (MOOCs) across disciplines. Several programming assignments are typically attempted by the students as a part of this course, which are evaluated and graded against pre-defined test-cases. Given the importance of programming education and the difficulty of providing relevant feedback for the massive number of students, there has been increasing interest in automated program repair techniques for providing automated feedback to student assignments (\APRS)~\cite{yi2017feasibility, clara, sarfgen, refactory, gulwani-fse16, singh-pldi13}.

\paragraph*{Existing approaches and their limitations}
Table \ref{tab:tools} provides a summary of state-of-the-art \APRS works, and compares them with our approach Verifix. The repair rate of the state-of-the-art \APRS techniques~\cite{clara, sarfgen, refactory} is astonishingly high, around 90\%. However, these works make certain assumptions such as the presence of tests. 


Many student assignment feedback generation approaches assume the existence of a complete set of high quality test-cases to validate their repairs. Over-fitting the repair to an incomplete specification is a well known problem of APR tools~\cite{LPR19,Qi2015}. 
Prior studies have shown that trivial repairs such as functionality deletion alone can achieve \textasciitilde50\% accuracy on buggy student programs given a weak oracle~\cite{chhatbar2020macer}.
Generating complex incorrect feedback that merely passes all the tests can potentially confuse novice students more than expert programmers.
Novice students when provided with incorrect/partial repair feedback that merely pass more tests, have been shown to struggle more, as compared to expert programmers given the same feedback~\cite{yi2017feasibility}. Hence, we suggest that the feedback given to novice students needs rigorous quality assurance, whenever possible. 

In a related vein, some approaches, in particular recent ones~\cite{clara, sarfgen}, assume the existence of multiple reference programs.
This assumption is made to overcome the difficulty of generating feedback when the Control-Flow Graph (CFG) structure of the student program is different from the instructor provided reference program.\footnote{SarfGen~\cite{sarfgen} and Clara~\cite{clara} require that the control-flow structure of student and reference programs should be exactly the same. Clara~\cite{clara} demands aligned variables to be evaluated into the same sequence of values at runtime, which often does not hold for early stage programs.} The problem is that the existing approaches collect multiple reference programs from student submissions that pass all tests, again without verifying their correctness. It is well-known that verifying the equivalence of two program is challenging~\cite{churchill2019semantic}.


\paragraph*{Insight}
Many of the aforementioned problems of the existing \APRS techniques can be addressed with verified repair. We can assume only the presence of one reference assignment, which is always available in educational settings and can be given by an instructor. We then create a verifiably correct repair of the student assignment. In other words, the repaired student assignment will be semantically equivalent to the reference assignment given by the instructor.
In terms of workflow, the repair engine indicates when it can generate a verified repair as feedback, and when it does, the students can receive a feedback which is guaranteed to be correct. In other words, we can have greater confidence or trust on the feedback generated by the repair tool. Furthermore, student programs that are verified to be correct after repair can be used as additional trustworthy reference programs in future . 

\paragraph*{Contribution: Verified repair}
In this work, we propose a general approach to verified repair, and show that verified repair can be performed with a single reference program. Verified repair engenders greater trust in the output of the automatic repair tool, which has been identified to be a key hindrance in deployment of automated program repair \cite{ryan2019trust}.
We show that verified repair is feasible and achievable in a reasonable time scale (less than 30 seconds) for student programming assignments of a large public university. This shows the promise of using verified repair to generate high confidence {\em live} feedback in programming pedagogy settings. 
To the best of our knowledge, ours is the first work to espouse verified repair for general purpose programming education.
The only previous attempt on verified repair~\cite{gulwani-fse16} is tightly tied to a specific structure of programs implementing dynamic programming.


\paragraph*{Repair tool: Verifix} 
We build our verified-repair technique by extending the existing program equivalence checking technique. Although automatically proving the equivalence between two programs remains challenging (mainly due to the difficulty of automatically finding loop invariants), we found that student programs are in many cases amenable for equivalence checking. This is because there is usually a reference program whose structure is similar to the student program, as shown in earlier works~\cite{clara,sarfgen}. Exploiting this, Verifix 
produces a verified repair. Note that, Verifix peforms repair and equivalence checking at once. 
More concretely, Verifix aligns the incorrect student program with the reference program into an aligned automaton, derives alignment relation to relate the variable names of the two programs, and suggests repairs for the code captured by the edges of the aligned automaton via Maximum Satisfiability-Modulo-Theories (MaxSMT) solving. We use MaxSMT to find a minimal repair. 
Our approach can generate a program behaviourally equivalent to the reference program while preserving the original control-flow of the student program as much as possible. This leads to smaller patches/feedback which we believe are easier to comprehend, in general.
We evaluate our approach on student programming submissions curated from a widely used intelligent tutoring system. Our approach Verifix produces small-sized verified patches as feedback, which, whenever available, can be used by struggling students with high confidence.


\begin{figure*}[!ht]
  \def\figwidth{0.48\textwidth}
    \begin{subfigure}[b]{\figwidth}
    \centering
    \includegraphics[width=\columnwidth]{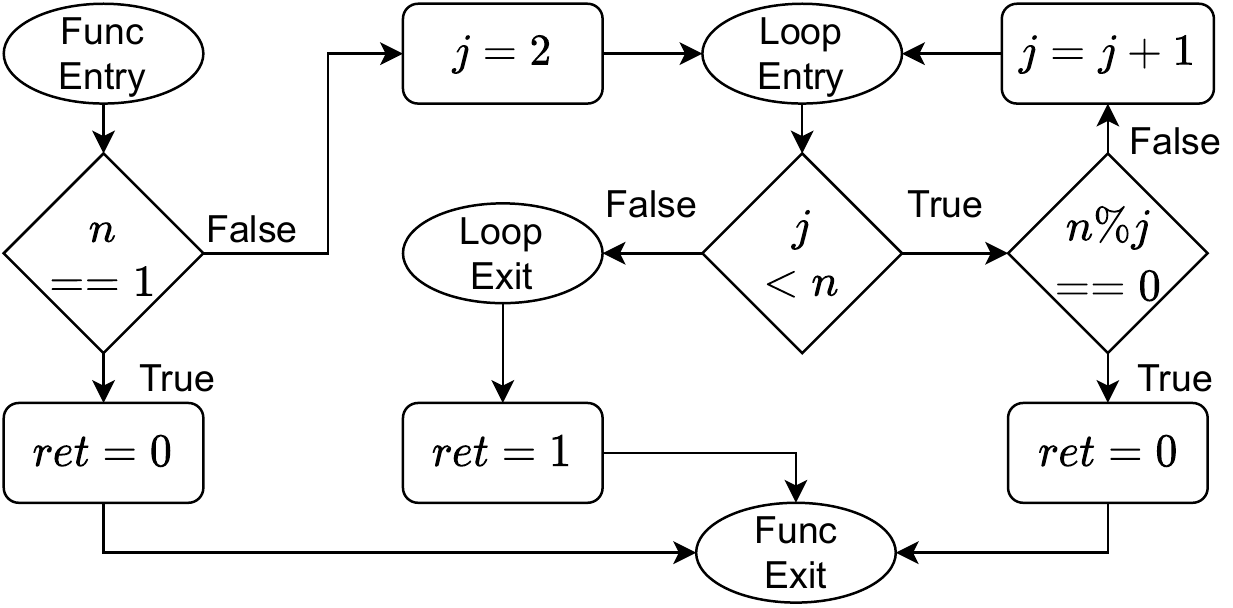}
    \caption{CFG of the reference program in Fig~\ref{fig:example-ref}}
    \label{fig:cfg-reference}
    \end{subfigure}
    \hfill
    \begin{subfigure}[b]{\figwidth}
    \centering
    \includegraphics[width=\columnwidth]{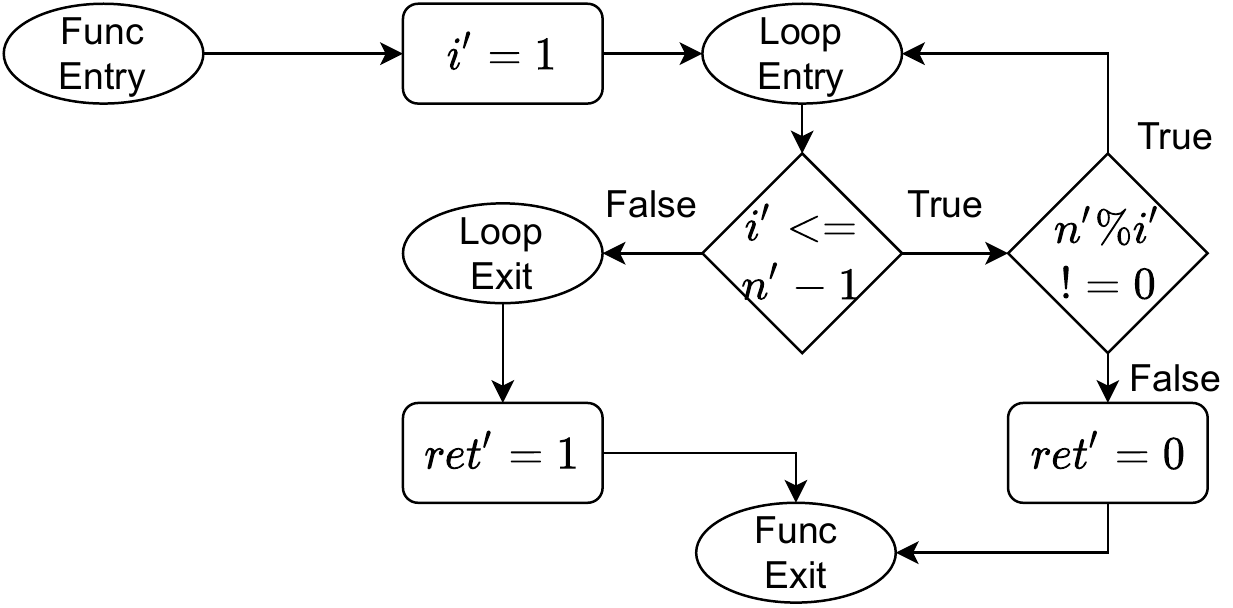}
    \caption{CFG of the incorrect student program in Fig~\ref{fig:example-incorrect}}
    \label{fig:cfg-student}
    \end{subfigure}    
  %
    \caption{Control Flow Graph (CFG) of the reference and incorrect program listed in Fig~\ref{fig:example}. Incorrect program CFG in Fig~\ref{fig:cfg-student} differs from reference program CFG in Fig~\ref{fig:cfg-reference} due to a missing return node. Existing tools like Clara~\cite{clara}, Sarfgen~\cite{sarfgen} cannot repair the incorrect program.}
    \label{fig:cfg}
\end{figure*}

\begin{figure*}[!ht]
    \begin{subfigure}[b]{.39\textwidth}
        \includegraphics[width=\columnwidth]{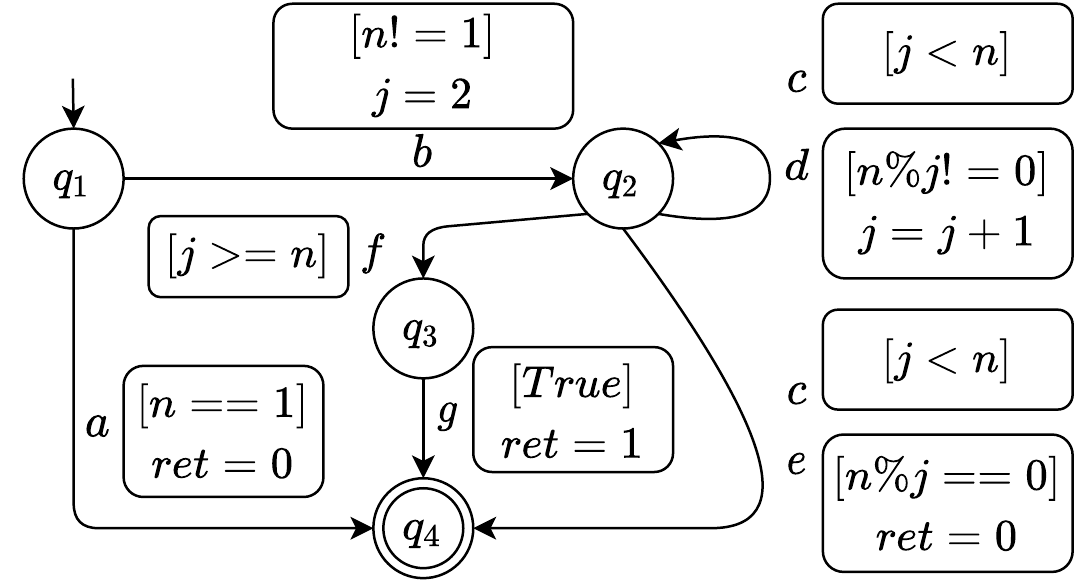}
        \caption{Reference CFA $\mathcal{A}_R$}
        \label{fig:cfa-reference}
    \end{subfigure}
    %
    \begin{subfigure}[b]{.39\textwidth}
        \includegraphics[width=\columnwidth]{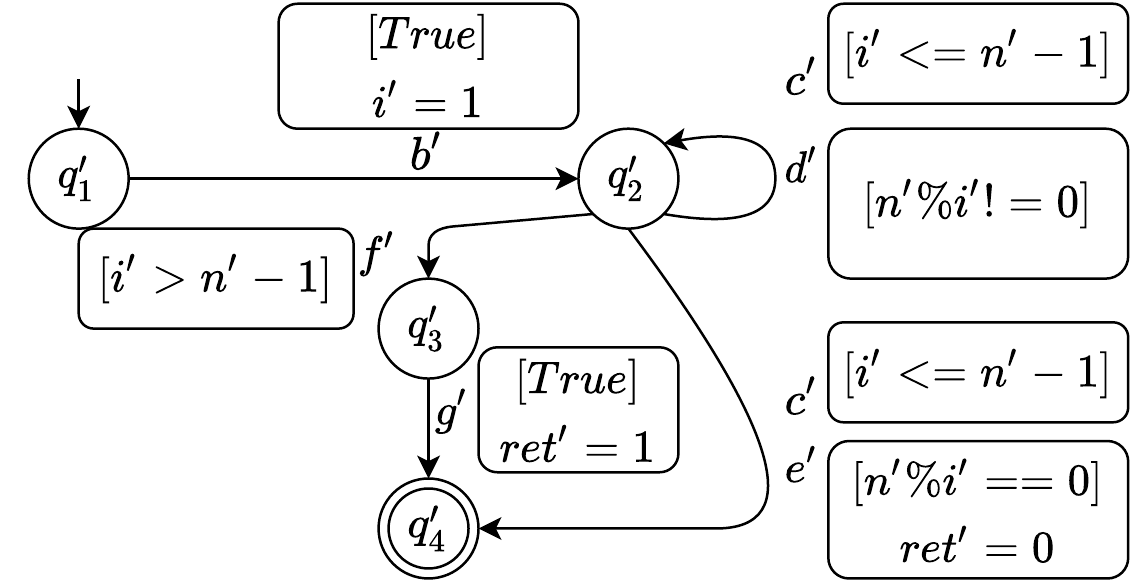}
        \caption{Student's CFA $\mathcal{A}_S$}
        \label{fig:cfa-student}
    \end{subfigure}    
    %
    \begin{subfigure}[b]{.2\textwidth}
        \centering
        \includegraphics[width=\columnwidth]{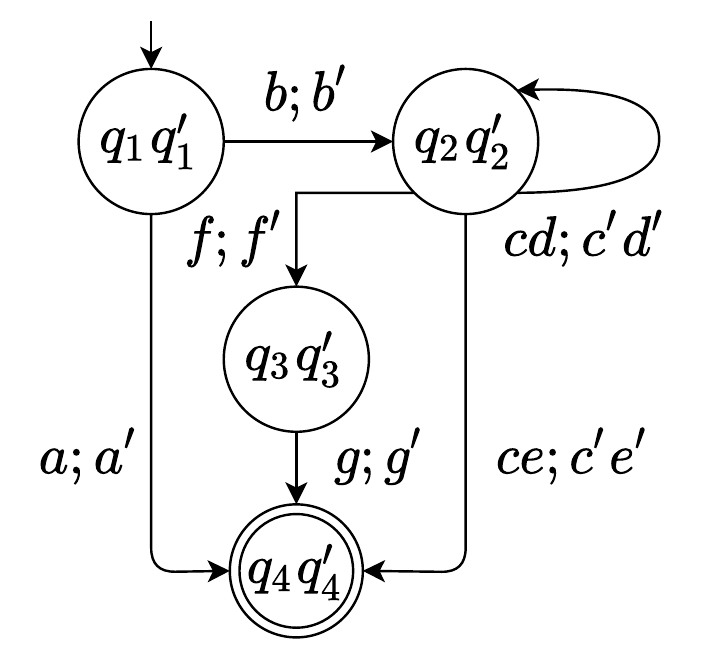}
        \caption{Aligned CFA $\mathcal{A}_F$}
        \label{fig:cfa-aligned}
    \end{subfigure}    
    \caption{Control Flow Automata (CFA) of the reference and incorrect program listed in Fig~\ref{fig:example}.
    CFA $\mathcal{A}_R$ of reference program in Fig~\ref{fig:cfa-reference} is structurally aligned with CFA $\mathcal{A}_S$ of student program in Fig~\ref{fig:cfa-student} to obtain an aligned CFA $\mathcal{A}_F$ in Fig~\ref{fig:cfa-aligned}.}
    \label{fig:cfa}
\end{figure*}



\section{Overview}
\label{sec:overview}

Consider a simple programming assignment for checking whether a given number $n$ is a prime number. 
Figure~\ref{fig:example-ref} shows a reference implementation prepared by an instructor, and Figure~\ref{fig:example-incorrect} shows an incorrect program submitted by students. 

\paragraph*{Limitations of the Existing Approaches}
The state-of-the-art approaches such as Clara~\cite{clara} and Sarfgen~\cite{sarfgen} make the same-control-flow assumption described as follows. 

\begin{assumption}
To perform a repair, a given incorrect program and its reference implementation should have the same control-flow structure.~\!\!\footnote{For the definitions of control-flow structure used in Clara and Sarfgen, refer to Definition~4.1 of \cite{clara} and Definition~3.1 of \cite{sarfgen}} 
\end{assumption}

Notice that this assumption does not hold for the incorrect program shown in Figure~\ref{fig:example-incorrect}. While the reference program has an \p{if} statement whose successors are either the \p{for} loop in lines~4--8 or the end of the program, the incorrect program starts with a \p{while} loop without having a preceding \p{if} statement.
As a result, both Clara and Sarfgen fail to generate a repair.

A common approach that has been used to overcome this problem is to use multiple reference programs of diverse control-flow structures~\cite{clara,sarfgen,gulwani-fse16}. Since it would be labor-intensive for an instructor to prepare multiple reference implementations, recent works (e.g., \cite{clara, sarfgen}) gets around this problem by using student submissions. That is, a student submissions that pass all tests are added into a pool of reference implementations. 


\paragraph*{Our Approach}
We show how we address the aforementioned limitations. Essentially, we do not make the same control-structure assumption, by conducting repair with Control Flow Automata (CFA) where its nodes represent program locations and its edges represent guarded commands (see Figure~\ref{fig:cfa} as an example). Also, we extend the existing equivalence checking technique into a verified repair technique.
In the following, we show our repair algorithm works through the following three phases: the setup phase, the verification phase, and the repair phase. The last two steps occur simultaneously as explained in the following. 

\subsection{Setup Phase}

In the setup phase, we model the given reference and student programs as Control Flow Automata (CFA) with the nodes representing control-flow locations and the edges representing guarded commands, as shown in Figure~\ref{fig:cfa}. It is possible for the guard to be true, in which case the command is executed unconditionally. 
Following which we prepare multiple aligned automata $\mathcal{A}_F$ using a syntax-based alignment technique (see Section~\ref{sec:StructuralAlignment}).
Figure~\ref{fig:cfa} shows an example of $\mathcal{A}_F$ obtained on aligning the reference automata in Figure~\ref{fig:cfa-reference} and student automata in Figure~\ref{fig:cfa-student}.

The aligned automata $\mathcal{A}_F$ in Figure~\ref{fig:cfa} consists of an aligned control entry node $q_1q'_1$, control exit node $q_4q'_4$, a loop entry node $q_2q'_2$, and a loop exit node $q_3q'_3$.
The notation 
\[q_2q_2' \stackrel{cd;c'd'}{\longrightarrow} q_2q_2'
\]
denotes that $q_2 \stackrel{cd}{\longrightarrow} q_2$ is aligned with $q_2' \stackrel{c'd'}{\longrightarrow} q_2'$ where $cd$ represents a sequence of edges $c$ and $d$. Note that in the case of $q_1q_1'\stackrel{aa'}{\longrightarrow}q_4q_4'$, a new edge $a'$ is introduced to connect two aligned nodes, $q_1q_1'$ and $q_4q_4'$. In general, it is possible that more than one syntactic aligned automaton can be created from two given assignments, in which case we enumerate over each aligned automaton to find out a minimal repair. 

Given an aligned automaton, the verification of behavioral equivalence between two assignments can be performed by checking whether $q \sim q'$ (i.e., $q$ is bisimilar to $q'$) holds for each aligned nodes $q$ and $q'$, e.g.  in Figure~\ref{fig:cfa}(c), $q_i$ is aligned with $q'_i$ for $i \in \{1,2,3,4\}$. 

So, the core task we have is to prove $q \sim q'$ for a pair of aligned nodes $q$ and $q'$, and also to generate a repair if the proof fails. We annotate each node with a variable alignment predicate. 
Variable alignment predicate provides a bijective mapping between the variables of two programs, which is necessary to prove their semantic equivalence.  In our example, we obtain the following variable alignment predicate (automatically): $\{ret \leftrightarrow ret', n \leftrightarrow n', j \leftrightarrow i'\}$ where $ret$ is a special variable holding the return value of the function under verification/repair.

\subsection{Verification Phase}

We perform verification for all aligned automata $\mathcal{A}_F$.
If verification succeeds for $\mathcal{A}_F$ or its repaired variation, semantic equivalence between student and reference programs is guaranteed (see Theorem~\ref{th:soundness}).
Verification is performed inductively for individual edge, starting from the outgoing edges of the initial node of $\mathcal{A}_F$ ($aa'$ and $bb'$ for our Figure~\ref{fig:cfa}).

Consider the edge $q_1q_1' \stackrel{b;b'}{\longrightarrow} q_2q_2'$.
Given this edge, we should prove the following: when $q_1 \sim q'_1$ is assumed, $q_2 \sim q_2'$ holds after executing $b;b'$.
We achieve this by checking
\[
\varphi_{edge}^1: \phi_{q_1q_1'} \wedge \psi_r \wedge \psi_s^1 \wedge \neg \phi_{q_2q_2'}\] 
where $\phi_{q_1q_1'}$ and $\phi_{q_2q_2'}$ denote the variable alignment predicates at node $q_1q_1'$ and $q_2q_2'$, respectively.
\begin{center}
\begin{tabular}{ll}
$\phi_{q_1q'_1}$: &  $(ret_0 = ret'_0) \wedge (n_0 = n'_0) \wedge (j_0 = i'_0)$\\
$\phi_{q_2q'_2}$: &  $(ret_1 = ret'_1) \wedge (n_1 = n'_1) \wedge (j_1 = i'_1)$
\end{tabular}
\end{center}

Meanwhile, $\psi_r$ and $\psi_s^1$ denote the guarded commands of $b$ and $b'$, respectively, in a Single Static Assignment (SSA) form, where
\begin{center}
\begin{tabular}{llll}
$\psi_r$: &  $(n_0 \neq 1 \implies j_1 = 2)$ & $\wedge$ & $(\neg (n_0 \neq 1) \implies j_1 = j_0)$\\
$\psi_s^1$: &  ($True \implies i_1' = 1)$ & $\wedge$ & $(\neg True \implies i_1' = i'_0)$
\end{tabular}
\end{center}
If $\varphi^1_{edge}$ is satisfiable, then $q_2 \sim q_2'$ does not hold, indicating verification failure.
We check the satisfiability of $\varphi^1_{edge}$ using an off-the-shelf SMT solver, Z3~\cite{z3}. 


\begin{table}[t]
    \centering \small
    \begin{tabular}{c|ll|ll}
        Block & \multicolumn{2}{c|}{Student Transition} & \multicolumn{2}{c}{Repaired Transition} \\
        \hline
        $a'$ & $\emptyset$ & & $[n'==1]$ & $ret'=0$ \\
        $b'$ & $[True]$ & $i'=1$ & $[n'!=1]$ & $i'=2$ \\
        $c'$ & $[i' <= n'-1]$ & & $[i' <= n'-1]$ &  \\
        $d'$ & $[n' \% i'!=0]$ &  & $[n' \% i'!=0]$ & $i'=i'+1$ \\
        $e'$ & $[n' \% i'==0]$ & & $[n'\%i' == 0]$ & $ret'=0$ \\
        $f'$ & $[i'>n'-1]$ & & $[i' > n'-1]$ &  \\  
        $g'$ & $[True]$ & $ret'=1$ & $[True]$ & $ret'=1$ \\  
    \end{tabular}

    \caption{Incorrect student blocks and their corresponding repairs generated by Verifix, after multiple rounds of edge verification-repair of Figure~\ref{fig:cfa} aligned automaton.}
    \label{tab:repair_blocks}
    \vspace{-15pt}
\end{table}

\subsection{Repair Phase}

For our running example, the SMT solver Z3 finds that $\varphi_{edge}^1$ is satisfiable under a certain assignment $\phi_{ce}^1$ which is 
\[
\phi_{ce}^1: n_0 = n'_0 = 1, j_0 = i'_0 = 0
\]
We call this generated satisfying assignment $\phi_{ce}^1$ as a {\em counter-example},  since it is a counter-example to $q_2 \sim q_2'$ holding.
Using this counter-example $\phi_{ce}^1$,
we perform a repair based on  counter-example-guided inductive synthesis or CEGIS strategy~\cite{cegis}.
Following CEGIS strategy, we look for a repair of $\psi_s^1$ which rules out the counter-example $\phi_{ce}^1$.
Verifix returns  two potential repair candidates.
\begin{center}
\begin{tabular}{llll}
$\psi_s^2$: &  $(False \implies i_1' = 1)$ & $\wedge$ & $(\neg False \implies i_1' = i'_0)$\\
$\psi_s^3$: &  $(n'_0 \neq 1 \implies i_1' = 1)$ & $\wedge$ & $(\neg(n'_0 \neq 1) \implies i_1' = i'_0)$
\end{tabular}
\end{center}

Then, we repeat edge verification $\varphi_{edge}^1$ step with each of the obtained repair candidates.
In our example, verification attempt fails again for both repair candidates, that is 
$\varphi_{edge}^1$ is satisfiable. 
and a new counter-example $\phi_{ce}^2$ is obtained. 
\[
\phi_{ce}^2: n_0 = n'_0 = 2, i_0 = i'_0 = 0
\]
By considering both $\phi_{ce}^1$ and $\phi_{ce}^2$, Verifix returns a new repair candidate $\psi_s^4$,
\begin{center}
\begin{tabular}{llll}
$\psi_s^4$: &  $(n'_0 \neq 1 \implies i_1' = 2)$ & $\wedge$ & $(\neg(n'_0 \neq 1) \implies i_1' = i'_0)$
\end{tabular}
\end{center}

This repair candidate $\psi_s^4$ rules out both the counter-examples seen so far, and no further 
satisfying assignments of $\varphi_{edge}^1$ are found. This completes the verification and repair, thereby repairing the edge $b'$ in Figure \ref{fig:cfa}(b). 
Table~\ref{tab:repair_blocks} summarizes the buggy student automata $\mathcal{A}_S$ edges and the corresponding repairs generated by our repair tool Verifix.


Verifix searches for a minimal repair, for each aligned edge under consideration.
Informally, a minimal repair modifies the minimum number of expressions, and a more formal description is available in Theorem~\ref{th:minimality}.
Minimal repair is achieved by reducing our repair search into a partial MaxSMT problem.
Intuitively, a minimal repair should preserve the maximum number of the original expressions. 
Given the original guarded action of the student program, $\psi_s^1$, we first replace each of the original expressions of $\psi_s^1$ with a unique hole.
For example, $[True] \implies (i'_1 = 1)$ is transformed into $[h_1] \implies (i'_1 = h_2)$ where each hole $h_i$ can be filled in with a patch candidate expression chosen from the implementation space of $h_i$.
Clearly, the implementation space of $h_i$ also includes the original expression.
Verifix retains the original expressions for as many holes as possible, by associating a higher weight penalty with the original expressions over other replacement expressions. 
Such a choice of weights ensures that the original expressions are favored by the partial MaxSMT solver.
Further details are provided in Section~\ref{sec:repair}.

\section{Program Model}
\label{sec:model}
Prior to explaining our alignment and verification-repair procedures, we introduce the key structures used to model programs.

\subsubsection*{Abstract Syntax Tree (AST)}
An Abstract Syntax Tree (AST) consists of a set of nodes representing the abstract programming constructs. With the tree hierarchy, or edges, representing the relative ordering between the appearance of these constructs. 
We extend the standard AST with special labels for two node types: \textit{Func-Entry} and \textit{Loop-Entry}.
Each AST consists of a root node corresponding to a function definition, which is labelled as a function-entry node.
Similarly, every loop construct in the AST is labelled as a loop-entry node.

The AST for motivating example shown in Figure~\ref{fig:example} consists of two labelled nodes: a \textit{Func-Entry} node $q_1$ which maps to the \textit{check\_prime} function definition and a \textit{Loop-Entry} node $q_2$ which maps to the for-loop construct. We note that some existing APR techniques for programming assignments, like ITSP~\cite{yi2017feasibility} which uses GenProg~\cite{genprog}, operate on program ASTs directly. 

\subsubsection*{Control Flow Graph (CFG)}
Existing state-of-art APR techniques like Clara~\cite{clara} and SarfGen~\cite{sarfgen} operate at the level of CFG, whose nodes are basic blocks and edges denote control transfer.
We extend the standard CFG by introducing 4 types of special labelled nodes: \{\textit{Func-Entry}, \textit{Loop-Entry}, \textit{Func-Exit}, and \textit{Loop-Exit}\}; denoting the program states when control enters a function or a loop, and when control exits a function or a loop, respectively.
The \textit{Func-Entry} and \textit{Loop-Entry} CFG nodes correspond with control entering AST nodes of the same type. 
The \textit{Func-Exit} and \textit{Loop-Exit} CFG nodes correspond with the program state after control visits the last child of \textit{Func-Entry} and \textit{Loop-Entry} AST node, respectively.
These \textit{Func-Exit} and \textit{Loop-Exit} program states can also be reached by altering the control-flow using \textit{return} and \textit{break} statements, respectively.

Figures~\ref{fig:cfg-reference} and~\ref{fig:cfg-student} depict the CFG of the reference and student program in Figures~\ref{fig:example-ref} and~\ref{fig:example-incorrect}, respectively.
These CFGs contain four special nodes denoting \textit{Func-Entry} ($q_1/q'_1$), \textit{Loop-Entry} ($q_2/q'_2$), \textit{Loop-Exit} ($q_3/q'_3$), and \textit{Func-Exit} ($q_4/q'_4$) program states.

\subsubsection*{Control Flow Automata (CFA)}

Our tool Verifix operates at the level of the control flow automaton (CFA), often used by model-checking and verification communities \cite{lazy02}. The CFA is essentially the CFG, with code statements labeling the edges of CFA, instead of code statements labeling nodes as in CFG. 
The nodes of our CFA are annotated with the node types mentioned earlier: \textit{Func-Entry}, \textit{Loop-Entry}, \textit{Func-Exit}, and \textit{Loop-Exit}.
The edges of our CFA are constructed by choosing all possible code transitions between the program states in CFG. Depending on the reason for control-flow transition, these edges can be of three types: \textit{normal}, \textit{return} or \textit{break}.
Figures~\ref{fig:cfa-reference} and~\ref{fig:cfa-student} depict the CFA modeled using the  reference and student CFG in Figures~\ref{fig:cfg-reference} and~\ref{fig:cfg-student}, respectively.

We provide our precise definition of CFA in the following.

\begin{definition}[Control Flow Automata]
\label{def:cfa}
A Control Flow Automata (CFA) is a tuple of the form $\langle V, E, v^0, v^t, \Omega, \mathcal{\Uppsi}, Var\rangle$, where:
\begin{itemize}
\item $V$ : is a finite set of vertices (or nodes) of the automata, representing function and loop entry/exit program states,
\item $E \subseteq V \times V$, is a finite set of edges of the automata representing normal, break, and return transitions between program states,
\item $v^0$ : is the initial node representing function entry state,
\item $v^t$ : is the terminal node representing function exit state,
\item $\Omega$ : $\{u \leftrightarrow v \mid \forall u \in V, \exists v \in V\}$, for each function/loop entry node, maintains a mapping to the corresponding exit node,
\item $\Uppsi$ is mapping from edge $e$ to 
$\psi_e$ for all edges $e$, where $\psi_e$ is the set of guarded actions labeling $e$
\item $Var$, is the set of variables used in $\bigcup_{e} \psi_e$
\end{itemize}
\end{definition}
For edge $e$ in the CFA, $\psi_e$ is thus the code statements labeling $e$.

\section{Aligned Automata}
\label{sec:method}

\begin{figure}[t]
  \def\figwidth{0.48\columnwidth}
    \begin{subfigure}[b]{\figwidth}
    \centering
    \includegraphics[width=.7\columnwidth]{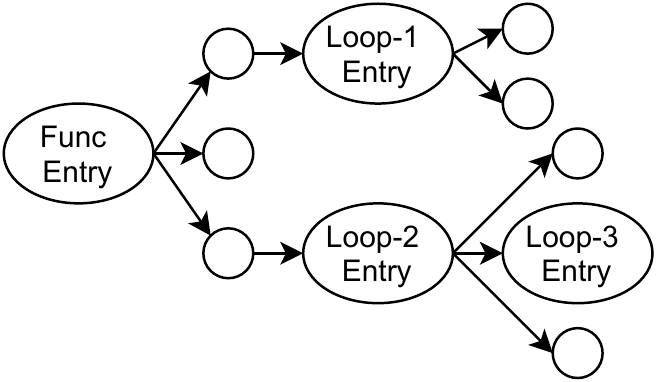}
    \caption{Example $AST$ with labelled and unlabelled nodes.}
    \label{fig:ast-before}
    \end{subfigure}
    \hfill
    \begin{subfigure}[b]{\figwidth}
    \centering
    \includegraphics[width=.7\columnwidth]{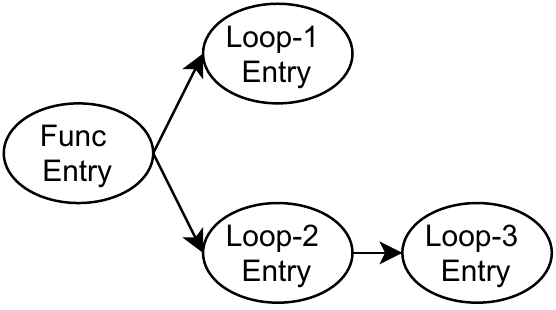}
    \caption{Example $AST^L$ after deletion of unlabelled nodes.}
    \label{fig:ast-after}
    \end{subfigure}    
  %
    \caption{Example demonstrating Abstract Syntax Tree (AST) transformation to retain nodes labelled as function and loop entry.}
    \label{fig:ast}
\end{figure}

\begin{figure}[t]
  \def\figwidth{0.48\columnwidth}
    \begin{subfigure}[b]{\figwidth}
    \centering
    \includegraphics[width=.5\columnwidth]{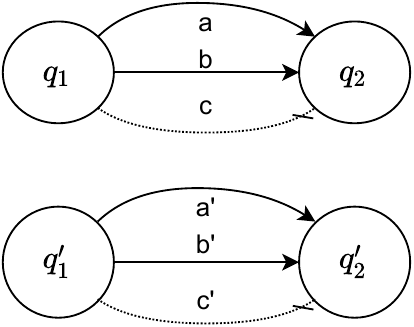}
    \caption{Example $\mathcal{A_R}$ and $\mathcal{A_S}$.}
    \label{fig:edge-before}
    \end{subfigure}
    %
    \begin{subfigure}[b]{\figwidth}
    \centering
    \includegraphics[width=.5\columnwidth]{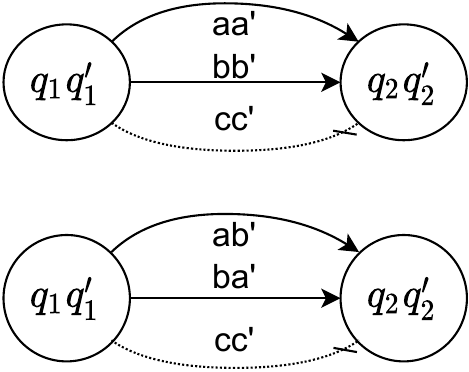}
    \caption{Example $\mathcal{A_F}$ after alignment.}
    \label{fig:edge-after}
    \end{subfigure}    
  %
    \caption{Example demonstrating edge alignment. Given node alignment $V:\{q_1q'_1, q_2q'_2\}$, the edges are aligned based on type. The single \textit{break} transitions $c$ and $c'$ are aligned with each other, while the multiple \textit{normal} edges are aligned combinatorially to produce two unique aligned automata.}
    \label{fig:edge}
\end{figure}

Our methodology for repairing incorrect student programs relies on constructing an aligned automaton  $\mathcal{A}_F$ from the given student automaton $\mathcal{A}_S$ and the reference automaton $\mathcal{A}_R$.
The construction of the automaton $\mathcal{A}_F$ consists of following steps:
(i) Model the student and reference programs as Control Flow Automata $\mathcal{A}_S$ and $\mathcal{A}_R$,
(ii) the structural alignment of $\mathcal{A}_S$ and $\mathcal{A}_R$, 
(iii) the inference of the variable alignment predicates.

\subsection{Structurally Aligning $\mathcal{A}_S$ and $\mathcal{A}_R$} 
\label{sec:StructuralAlignment}

For constructing aligned automata $\mathcal{A_F}$, we first construct a node alignment between the program states of reference and student automata. Followed by aligning the transition edges between the reference and student program states.

\paragraph*{Node Alignment}
Given two Control Flow Automata $\mathcal{A_S}$ and $\mathcal{A_R}$, and their corresponding Abstract Syntax Trees $AST_S$ and $AST_R$ for student and reference program, respectively, we construct node alignment $V : V_S \leftrightarrow V_R$ as follows.
\begin{enumerate}
    \item Delete all unlabelled nodes from $AST_S$ and $AST_R$ to obtain $AST^L_S$ and $AST^L_R$, respectively. An $AST^L$ consists of only \textit{Func-Entry} and \textit{Loop-Entry} labelled nodes.
    \item If the syntactic tree structure of $AST^L_S$ and $AST^L_R$ is same, align each node of $AST^L_S$ with $AST^L_R$ and add to $V$. This step aligns the \textit{Func-Entry} and \textit{Loop-Entry} nodes of $\mathcal{A_S}$ and $\mathcal{A_R}$.
    \item For each pair of entry nodes (either \textit{Func-Entry} or \textit{Loop-Entry}) that are aligned with each other, their corresponding exit nodes (either \textit{Func-Exit} or \textit{Loop-Exit}) are aligned with each other.
\end{enumerate}

For constructing node alignment $V$, we first align the labelled nodes of student and reference Abstract Syntax Tree (AST). 
The labelled AST nodes can be of two types: \textit{Func-Entry} and \textit{Loop-Entry}. These labels are same as those in $\mathcal{A}_S$ and $\mathcal{A}_R$, but we take advantage of the tree structure in the AST. 
Figure~\ref{fig:ast} demonstrates unlabelled $AST$ node deletion in step-1 through an example, after which only the \textit{Func-Entry} and \textit{Loop-Entry} labelled nodes are retained.
For the reference program (respectively student program) listed in Figure~\ref{fig:example}, the labelled $AST^L_R$ (resp. $AST^L_S$) consists of two nodes $q_1 \rightarrow q_2$ (resp. $q'_1 \rightarrow q'_2$). 
Since both the $AST^L$ trees are structurally similar, the node alignment $V$ consists of $\{q_1q'_1, q_2q'_2\}$ after step-2 of node alignment, denoting the \textit{Func-Entry} and the \textit{Loop-Entry} aligned nodes.

The step-3 of node-alignment finally aligns the function and loop exit nodes. Given the student and reference automata in Figure~\ref{fig:cfa}, $q_4$, which is the \textit{Func-Exit} node corresponding to $q_1$, is aligned with $q_4'$, which is the \textit{Func-Exit} node corresponding to $q_1'$. Similary, the \textit{Loop-Exit} nodes $q_3$ and $q'_3$ are aligned, since their corresponding \textit{Loop-Entry} nodes $q_2$ and $q'_2$ were aligned in step-2.
 

 
The node alignment constructed thus, if successful, will lead to a bijective mapping from nodes of $\mathcal{A_S}$ to nodes of $\mathcal{A_R}$. 
Intuitively, we cannot repair student programs that have a different function/looping structure from the reference program.
Note that the existing state-of-the-art approaches~\cite{clara,sarfgen} require identical number of edge transitions, in addition to an identical number of function/loop nodes, unlike in our approach.
Referring to Figure~\ref{fig:cfa}, we obtain the nodes $\{q_1q'_1, q_2q'_2, q_3q'_3, q_4q'_4\}$ of aligned automata $\mathcal{A_{F}}$ after node-alignment on student automata $\mathcal{A_S}$ and reference automata $\mathcal{A_R}$.


\paragraph*{Edge Alignment}
Given two CFAs $\mathcal{A_S}$ and $\mathcal{A_R}$, and their corresponding node alignment $V: V_S \leftrightarrow V_R$, we construct an aligned CFA $\mathcal{A_F}$ by aligning the edges of $\mathcal{A_S}$ and $\mathcal{A_R}$. Suppose that $u_S \leftrightarrow u_R$ (i.e., node $u_S$ in $\mathcal{A_S}$ is aligned with $u_R$ in $\mathcal{A_R}$) and $v_S \leftrightarrow v_R$. 
For each edge of type $t \in \{break, return, normal\}$, we treat the following three cases differently.

\begin{enumerate}
    \item $\mathcal{A_S}$ has only one edge from $u_S$ to $v_S$ of type $t$, and $\mathcal{A_R}$ has only one edge from $u_R$ to $v_R$ of the same type $t$.
    \item Only $\mathcal{A_R}$ has an edge from $u_R$ to $v_R$ of type $t$, while $\mathcal{A_S}$ has no edge from $u_S$ to $v_S$ of type $t$.
    \item None of the above matches, and $\mathcal{A_S}$ (or $\mathcal{A_R}$) has multiple edges from $u_S$ to $v_S$ (or from $u_R$ to $v_R$) of type $t$.
\end{enumerate}

In the first case, we simply align the matching edges. For example, in Figure~\ref{fig:cfa}, $\mathcal{A_R}$ contains only one \textit{normal} edge $b$ between $q_1$ and $q_2$ and $\mathcal{A_S}$ contains only one \textit{normal} edge $b'$ between $q_1'$ and $q_2'$. Hence, the aligned CFA $\mathcal{A_F}$ has an edge $b;b'$ as shown in Figure~\ref{fig:cfa-aligned}. 
An example of the second case is shown with the two nodes, $q_1q_1'$ and $q_4q_4'$, of $\mathcal{A_F}$. While $\mathcal{A_R}$ has one edge $a$ between $q_1$ and $q_4$, $\mathcal{A_S}$ has no edge between $q_1'$ and $q_4'$. In this case, we insert an edge $a;a'$ to $\mathcal{A_F}$ where $a'$ has an empty guarded action. 

Lastly, in the third case, there exist several possible edge alignments, of the order of $\binom{M}{N} \times N!$, where $M$ is the number of edges from $u_R \rightarrow v_R$ and $N$ is the number of edges from $u_S \rightarrow v_S$.
Fig~\ref{fig:edge} demonstrate this case through an example, resulting in two possible edge alignments.
We choose the edge alignment which maximizes the number of verified-equivalent edges in the resultant aligned automaton $\mathcal{A_{F}}$. The resultant aligned automaton $\mathcal{A_{F}}$ of our running example is shown Fig. \ref{fig:cfa}. The formal structure of aligned automaton appears in the following.

\begin{definition}[Aligned automaton]
\label{def:paa}
The automaton $\mathcal{A_{F}}$ that results from aligning the automata $\mathcal{A}_S$ and  $\mathcal{A}_R$  is a tuple of the form $\langle V, E, v^0, v^t, \Omega, \mathcal{\Uppsi}, Pred\rangle$, where:
\begin{itemize}
\item $V : V_S \leftrightarrow V_R$, is a finite set of one-to-one bijective mappings between the nodes of the automata $\mathcal{A}_S$ and $\mathcal{A}_R$, 
\item $E \subseteq V \times V$, is a finite set of edges representing normal, break, and return transitions between the aligned nodes,
\item $v^0 : v^0_S \leftrightarrow v^0_R$, where $v^0_S$ and $v^0_R$ are the initial function entry nodes of the automata  $\mathcal{A}_S$ and  $\mathcal{A}_R$ respectively, 
\item $v^t$ : $v^t_S \leftrightarrow v^t_R$, where $v^t_S$ and $v^t_R$ are the final function exit nodes of the automata  $\mathcal{A}_S$ and  $\mathcal{A}_R$ respectively
\item $\Omega$ : $\{u \leftrightarrow v \mid \forall u \in V, \exists v \in V\}$, for each function/loop entry node, maintains a mapping to the corresponding exit node,
\item $\Uppsi$ is the mapping from edge $e$ to $\psi_e$ for all edges $e$, where $\psi_e = \psi_s \cup \psi_r$, and $\psi_s$, $\psi_r$ are the set of guarded actions at the aligned edges $e_s$ and $e_r$ of the automata $\mathcal{A}_S$ and $\mathcal{A}_R$ respectively.
\item $Pred : Var_{S} \leftrightarrow Var_{R}$, denoting variable alignment, is a bijective mapping between variables of $\mathcal{A}_S$ and $\mathcal{A}_R$, 


\end{itemize}
\end{definition} 
Aligned automata augment the classical Control Flow Automata with variable alignment predicates for capturing mismatches (semantic variations) between the behavior of the student automaton and the corresponding reference automaton.

\subsection{Inferring Variable Alignment Predicates } \label{sec:VarAligning}

To infer alignment predicates  of $\mathcal{A}_F$, we use a syntactic approach based on variable-usage patterns similar to that of SarfGen\cite{sarfgen}.
Our approach for computing the mapping between two sets of variables proceeds as follows.

For each edge $e_i$ in $\mathcal{A}_F$ we collect the usage set for each variable $x/x'$ in the reference/student program, namely the sets $usage(x, e_i)$ and $usage(x', e_i)$. 
In case the student automaton has fewer variables than reference automaton ($|Var_{S}|<|Var_{R}|$), then fresh variables are defined in $Var_S$.
The goal is to find a variable alignment, a bijective mapping between $Var_{R}$ and $Var_{S}$, which minimizes the average distance between $usage(x, e_i)$ and $usage(x', e_i)$ for each $i \in [1, n]$, where $n$ is the number of edges in $\mathcal{A}_F$.
This is done by constructing a distance matrix $\mathcal{M}_{e_{i}}$ for each edge $e_i$ of size $|Var_R| \times |Var_S|$, where  
\[
\mathcal{M}_{e_i} (x, x') = \Delta ~ (usage(x, e_i), usage(x', e_i))
\]
Using the matrices $\mathcal{M}_{e_{1}},\ldots, \mathcal{M}_{e_{n}}$, we construct a global distance matrix $\mathcal{M}_g$ for the entire set of edges in  $\mathcal{A}_F$, where 
\[
\mathcal{M}_g(x, x') = \sum_{i = 1}^{n} \frac{\mathcal{M}_{e_{i}} (x, x')}{n}
\]
We then choose to align the variable $x$ in $R$ to the variable $x'$ in $S$, denoted as $x \leftrightarrow x'$, if the pair $(x, x')$ has the minimum average distance among all possible variable $y$ aligned with $x'$, that is among all variable alignment pairs $(y,x')$.

\section{Verification and Repair Algorithm}
\label{sec:VerifyAndRepair}



Once the aligned automaton $\mathcal{A}_F$ is constructed, we can initiate the repair process of the incorrect student program.
Note that a repaired version of the incorrect student program produced by our algorithm is guaranteed to be semantically equivalent to the given  structurally matched reference program. 
Our algorithm traverses the edges of the automaton $\mathcal{A}_F$ to perform edge verification  which basically checks the semantic equivalence between an edge of the student automaton and its corresponding edge of the reference automaton.\footnote{Our implementation performs a breadth-first search, while our algorithm is not restricted to a particular search strategy.}
In case the edge verification fails, we perform edge repair after which edge verification succeeds. 
While the existing approaches~\cite{clara,sarfgen} also similarly perform repair for aligned statements/expressions, the correctness of repair is not guaranteed unlike in our algorithm.   

We combine the edge verification and repair into a single step by extending the well-known SyGuS (syntax-guided synthesis) approach~\cite{sygus} which can be defined as follows:

\begin{definition}[SyGuS]
SyGuS consists of $\langle \varphi, T, S \rangle$ where $\varphi$ represents a correctness specification expressed assuming background theory $T$ and $S$ represents the space of possible implementations ($S$ is typically defined through a grammar). The goal of SyGuS is to find out an implementation that satisfies $\varphi$. 
\end{definition}

While in principle SyGuS can be directly used to perform repair, we have an additional non-functional requirement not considered in SyGuS---that is, we want to preserve the student program as much as possible for pedagogical purposes. 
To accommodate this additional requirement, we introduce our approach, SyGuR (syntax-guided repair), formulated as follows:

\begin{definition}[SyGuR]
\label{def:sygur}
Syntax-guided Repair or SyGuR consists of $\langle \varphi, T, S, impl_{o} \rangle$ where the first three components are identical with those of SyGuS, and $impl_{o} \in S$ represents the original implementation that should be repaired. The goal of SyGuR is to find out a repaired implementation $impl_{r} \in S$ that satisfies $\varphi$. In addition, differences between $impl_{o}$ and $impl_{r}$ should be minimal under a certain minimality criterion.
\end{definition}

\vspace{-3pt}
We realize SyGuR in the context of automated feedback generation for student programs. 
In this section, we present the two algorithmic steps we perform to conduct SyGuR: edge verification and edge repair.



\subsection{Edge Verification}
\label{sec:edge-verification}

In this section, we describe how we detect faulty expressions in the given incorrect student program. Recall that the edges of the  automaton $\mathcal{A}_F$ are constructed by aligning the edges of the student automaton $\mathcal{A}_S$ with the edges of the reference automaton $\mathcal{A}_R$.   Recall also that the edges of $\mathcal{A}_S$ can be  faulty while the edges of $\mathcal{A}_R$ are considered always as non-faulty.

Each edge $e: u \stackrel{\psi_s ; \psi_r}{\longrightarrow} v$ of $\mathcal{A}_F$ between nodes $u$ and $v$ asserts the following property:
\begin{equation}
\label{eq:simulation}
\{\phi_u\}\psi_s ; \psi_r \{\phi_v\}
\end{equation}
where $\phi_u$ and $\phi_v$ are the variable alignment predicates  at the source node $u$ and target node $v$ of the edge $e$ respectively, and $\psi_r$ and $\psi_s$ represent a list of guarded actions of the reference implementation and student implementation, respectively, expressed in a Single Static Assignment (SSA) form. For example, an original guarded action, \p{if (x>1) x++}, is converted into its SSA form, \m{((x_1 > 1) \implies x_2 = x_1+1) \land (\neg(x_1 > 1) \implies x_2 = x_1)}.
Note that $\psi_s$ and $\psi_r$ do not interfere with each other, since the variables used in $\psi_s$ and $\psi_r$ are disjoint from each other.
Also note that $\psi_r$ and $\psi_s$ do not contain a loop (that is, a single edge does not form a loop), and thus an infinite loop does not occur in the edge.

Edge verification succeeds if and only if property~(\ref{eq:simulation}) holds.
In SyGuR, property~(\ref{eq:simulation}) expresses a correctness specification $\varphi_e$ for edge $e$.
To check property~(\ref{eq:simulation}), we use an SMT solver by checking the satisfiability of the following formula:
\begin{equation} \label{eq:edge-verification}
\varphi_{e} = \phi_u \land \psi_s \land \psi_r \land \neg \phi_v
\end{equation}
The satisfiability of $\varphi_{e}$ indicates verification failure, or showing non-equivalence of two implementations along edge $e$. Conversely, the unsatisfiability of $\varphi_{e}$ indicates verification success. Note that there always exists a model $m$ that satisfies $\phi_u \land \psi_r \land \psi_s$ (this is because $\phi_u$ is not false, and the SSA forms of $\psi_r$ and $\psi_s$ are defined over disjoint variables), and verification succeeds only when for all such $m$, $\neg \phi_v$ does not hold. Intuitively, verification succeeds if and only if it is impossible for the post-condition $\phi_v$ to be false after executing $\psi_r$ and $\psi_s$ under the pre-condition $\phi_u$. 

As for background theories in the SMT solver, we use: LIA (linear integer arithmetic) for integer expressions, the theory of strings for modeling input/output stream, theory of uninterpreted functions to deal with function calls such as scanf/printf, and LRA (linear real arithmetic) to approximate floating-point expressions.

\subsection{Edge repair}
\label{sec:repair}

Once $\varphi_e$ is found to be satisfiable for an edge $e$ (which indicates that the edge verification fails), our goal is to repair edge $e$ by modifying the student implementation encoded in $\psi_s$.
Algorithm~\ref{alg:paa-repair} shows our edge repair algorithm based on the CEGIS (counter-example-guided inductive synthesis) strategy~\cite{cegis}. In step~1, edge verification is attempted, and verification failure results in a counter-example $\phi_{ce}$ that witnesses verification failure.
In the remaining part of the algorithm, we modify $\psi_s$ to $\psi'_s$ in a way that $\{\phi_{ce} \}\psi'_s ; \psi_r \{\phi_v\}$ holds.
If $\{\phi_u\}\psi'_s ; \psi_r \{\phi_v\}$ also happens to hold, edge repair is deemed as completed.
Otherwise, an SMT solver generates a new counter-example $\phi'_{ce}$, and our algorithm searches for $\psi''_s$ satisfying both $\{\phi_{ce} \}\psi''_s ; \psi_r \{\phi_v\}$ and $\{\phi'_{ce} \}\psi''_s ; \psi_r \{\phi_v\}$.
This process is repeated until either edge repair is successfully done or it fails. 
Edge repair can fail either because the search space is exhausted or timeout occurs. 

Let us first consider cases where $\psi_s$ and $\psi_r$ have the same number of guarded actions and all guarded actions have the same number of assignments. To ensure this requirement is met, we call function $Extend$ (see line~20 of Algorithm~\ref{alg:paa-repair} ) which will be described later.
Under the current assumption that $\psi_s$ and $\psi_r$ have the same number of guarded actions, $Extend(\psi_s, \psi_r)$ returns $\psi_s$, and thus, its return value $\psi_s^+$ equals $\psi_s$.

To repair guarded actions $\psi_s^+$, we replace each of the conditional expressions and the update expressions (RHS expressions) with a unique placeholder variable $h$.
This makes an effect of making holes in $\psi_s^+$, and filling in a hole for repair amounts to equating $h$ with a repair expression.
Function $RepairSketch$ of the algorithm performs this task of making holes in $\psi_s^+$ and returns $\psi_f$ defined as $\psi_s^+[e^{(i)} \mapsto h^{(i)}]$.
In this definition, notation $\mapsto$ denotes a substitution operator defined over all expressions $e^{(i)}$ appearing in $\psi_s^+$ and their corresponding placeholder variables $h^{(i)}$. 
In the following, we use \dquote{hole} to refer to a placeholder variable. 

In SyGuR (see Definition~\ref{def:sygur}), the expression space of the holes is defined by implementation space $S$.
Previous state-of-the-art works~\cite{clara, sarfgen} use the expressions of the reference program for repair (generated repairs are not verified in these works unlike in our approach), and we similarly define the implementation space of each hole as follows:

\begin{definition}[Implementation space of a hole] 
\label{RepairSynthesis} 
Let $C_s$ ($C_r$) and $U_s$ ($U_r$) be respectively the set of conditional and update expressions of $\psi_s^+$ ($\psi_r$).
Recall that $\psi_s^+$ ($\psi_r$) represents guarded actions of the student (reference) program.
When a conditional expression $e_c$ is replaced by a hole $h_c$, the implementation space of $h_c$ is defined as
$
C_s \mid C^{'}_r \mid true \mid false, 
$
where $C^{'}_r$ represents the set of conditional expressions appearing in $\psi_r$ with all variables of $C_r$ replaced with their aligned variables of the student program (see Section~\ref{sec:VarAligning} for variable alignment). 
Similarly, given an assignment $x = h_u$ where $h_u$ represents a hole for an update expression $e_u$, the implementation space of $h_u$ is defined as
$
U_s \mid U^{'}_r \mid x,
$
where $U^{'}_r$ represents the set of update expressions of $\psi_r$ with all variables of \,$U_r$ replaced with their aligned variables of the student program. 
The inclusion of an lhs variable $x$ in the implementation space is to allow assignment deletion---replacing $x=e_u$ with $x=x$ simulates assignment deletion.
\end{definition}

The repair synthesis process for some faulty expression on the edge $e_s$ relies on four factors:  the discovered counter-examples, the set of suspicious expressions in $\psi_s^+$, the set of reference expressions in $\psi_r$, and the inferred alignment predicates. These factors collectively determine the set of expressions on the edge $e_r$ that can be exploited to repair the buggy expressions on $e_s$. 
\begin{algorithm}[t]
\renewcommand{\algorithmicrequire}{\textbf{Input:}}
\renewcommand{\algorithmicensure}{\textbf{Output:}}
\small
\caption{Edge verification-repair}
\label{alg:paa-repair}
\begin{algorithmic}[1]
\Require Aligned $edge$ 
\Ensure Verified/Repaired $edge$

\State Let $\phi_u \equiv$ edge.sourceNode.invariants
\State Let $\phi_v \equiv$ edge.targetNode.invariants 
\State Let $\psi_r \equiv$ edge.label.reference
\State Let $\psi_s \equiv$ edge.label.student
\State $CEs \gets$ [] // List of counter-examples
\State $candidates \gets [\psi_s]$

\Repeat
    \State // Step 1: attempt for edge verification
    \For{each $\psi^i_s$ in $candidates$}
        \State Let $\varphi^i_{edge} \equiv \neg(\phi_u \wedge \psi_r \wedge \psi^i_s \implies \phi_v)$
        \If{$\not\models \varphi^i_{edge}$} // UNSAT 
            \State $edge.label.student \gets \psi^i_s$ // Update edge
            \State \Return \cmark \M// Verifiably correct
        \Else 
            \State $\phi^i_{ce} \models \varphi^i_{edge}$ // SAT
            \State $CEs \gets CEs \cdot \phi^i_{ce}$
        \EndIf
    
    \EndFor
    
    \State // Step 2: make holes in $\psi_s$
    \State Let $\psi_s^{+} \equiv Extend(\psi_s, \psi_r)$ 
    \State Let $\psi_f \equiv RepairSketch(\psi_s^{+})$ 
    
    \State // Step 3: define implementation space
    \State $\varphi_{hard} \gets []$;
    $\varphi_{soft} \gets []$
    \For{each $\phi^i_{ce}$ in $CEs$}
        \State $\varphi_{hard} \gets \varphi_{hard} \cdot (\phi^i_{ce} \wedge \psi_r \wedge \psi_f \wedge \phi_v)$
    \EndFor
    \For{each $hole, expr, weight$ in $RepairSpace(\psi_f, \psi_r, \psi_s)$}
        \State $\varphi_{soft} \gets \varphi_{soft} \cdot (hole = expr, weight)$
    \EndFor

    \State // Step 4: search for a repair    
    \If{$\not\models (\varphi_{hard}, \varphi_{soft})$} // UNSAT or UNKNOWN
        \State \Return \xmark \M// Repair Failure
    \Else
        \State // Update $candidates$ using a pMaxSMT solver
        \State // There can be multiple candidates
        \State $candidates \models (\varphi_{hard}, \varphi_{soft})$ 
    \EndIf

\Until{timeout}

\end{algorithmic}
\end{algorithm}

Recall that given a list of counter-examples $CE$, we search for a repair $\psi'_s$ that satisfies $\forall \phi_{ce} \in CE: \{\phi_{ce} \}\psi'_s ; \psi_r \{\phi_v\}$.
When searching for a repair, we preserve the expressions of the student program as much as possible for pedagogical reasons. 
We achieve this by conducting a search for a repair using a pMaxSMT (Partial MaxSMT) solver.
Note that an input to a pMaxSMT solver consists of (1) hard constraints which must be satisfied and (2) soft constraints all of which may not be satisfied.
Whenever a soft constraint $C$ is not satisfied, cost is increased by the weight associated with $C$, and a pMaxSMT solver searches for a model that minimizes the overall cost.
We pass the following formula to a pMaxSMT solver where hard constraints are underlined.
\begin{align*}
& \forall \phi_{ce} \in CE: (\underline{\phi_{ce}} \wedge \underline{\psi_r} \wedge \underline{\psi_f} \wedge \underline{\phi_v} \wedge \\
& \bigwedge_{(h^{(i)}, e^{(i)}, S\llbracket {h^{(i)}} \rrbracket) \in holes(\psi_f)} (h^{(i)} = e^{(i)} \wedge h^{(i)} \in S\llbracket {h^{(i)}} \rrbracket \backslash \{e^{(i)}\})) \numberthis \label{eq:pmaxsmt}
\end{align*}
where function $holes(\psi_f)$ returns a set of $(h^{(i)}, e^{(i)}, S\llbracket {h^{(i)}} \rrbracket)$ in which $h^{(i)}$ represents the placeholder variable appearing in $\psi_f$ (recall that $\psi_f$ is prepared by making holes in $\psi_s$), $e^{(i)}$ denotes the original expression of $h$ extracted from student program, and $S\llbracket {h^{(i)}} \rrbracket$ represents the implementation space of $h^{(i)}$.
Our soft constraints encode the property that each of the original expressions can be either preserved or replaced with an alternative expression in the implementation space. 
To preserve as many original expressions as possible, we assign a higher weight to $h^{(i)} = e^{(i)}$ than $h^{(i)} \in S\llbracket {h^{(i)}} \rrbracket \backslash \{e^{(i)}\}$.


\signpost{The Extend function}
Previously, we consider only the cases where $\psi_s$ and $\psi_r$ have the same number of guarded actions and all guarded actions have the same number of assignments.
To ensure this requirement, we invoke the $Extend$ function which performs the following.
First, if $\psi_s$ has a smaller number of guarded actions than $\psi_r$, then $\psi_s^+$ (the return value of $Extend$) should contain additional guarded actions, each of which uses the following template: $[False] \implies x = x$, where $x$ is constrained to be the variables of the student program.
Notice that these additional guards are initially deactivated to preserve the original semantics of the student program, but they can be activated whenever necessary during repair, since $False$ is replaced with a hole by $RepairSktech$. 
After this step, $Extend$ finds the guarded action of $\psi_r$ that has the maximum number of assignments.
Given this maximum number $M$, we check whether all guarded actions of $\psi_s$ (including additional guarded actions with the $False$ guard) also have $M$ assignments. 
Any guarded action that has a smaller number of assignments than $M$ is appended with additional assignments, $x = x$ where $x$ is constrained to be the variables of the student program.
This process makes sure that for each guarded action, the student program can have as many assignments as the reference program.

\subsection{Properties preserved by Verifix}

Once all the edges of the aligned automaton $\mathcal{A}_F$ are repaired and verified, it is straightforward to produce a repaired student automaton $\mathcal{A}'_S$ by copying repaired expressions from the automaton $\mathcal{A}_F$ to the automaton $\mathcal{A}_S$. In this section, we discuss several interesting properties of our repair algorithm, namely soundness, completeness, and minimality of generated repairs.
Some of their proofs are available in the supplementary material.




\begin{theorem}[Soundness]
\label{th:soundness}
For all program inputs, $\mathcal{A}'_S$ and $\mathcal{A}_R$ return the same program output.
\end{theorem}
%


Our edge repair algorithm (Algorithm~\ref{alg:paa-repair}) always returns a repaired edge as long as the underlying MaxSMT/pMaxSMT solver used in the algorithm is complete (that is, UNKNOWN is not returned). This can be stated as follows, using the concept of relative completeness~\cite{cook1978soundness}:

\begin{theorem}[Relative completeness of edge repair]
The completeness of Algorithm~\ref{alg:paa-repair} is relative to the completeness of the MaxSMT/pMaxSMT solver.
\end{theorem}

Meanwhile, the overall repair algorithm of Verifix is not complete.
If $\mathcal{A}_F$ is failed to be constructed, the repair process cannot be started. 
Theorem~\ref{th:completeness} identifies the conditions under which Verifix succeeds to generate a repair.
In Theorem~\ref{th:completeness}, we use the following definition of alignment consistency:

\begin{definition}[Alignment Consistency]
\label{def:consistency}
For each edge $e$ of $\mathcal{A}_F$  $\{\phi_u\}\psi_s ; \psi_r \{\phi_v\}$, modify $\psi_s$ into the $\psi'_s$ as follows:  $\psi'_s \equiv \psi_r[x_r^{(i)} \mapsto x_s^{(i)}]$ where $x_r^{(i)}$ denotes all reference-program variables appearing in $\psi_r$ and $x_s^{(i)}$ denotes student-program variables aligned with $x_r^{(i)}$. Repeat this for all edges of $\mathcal{A}_F$. Then, we say that $\mathcal{A}_F$ is alignment consistent when $\{\phi_u\}\psi'_s ; \psi_r \{\phi_v\}$ for all modified edges.
\end{definition}

$\mathcal{A}_F$ is alignment consistent only when the variable alignment predicates are such that a given student program can be verifiably repaired by edge-to-edge copy of the reference program (patch minimality is not considered).

\begin{theorem}[Relative completeness]
\label{th:completeness}
Our repair algorithm succeeds to generate a repair, under the following assumptions:
\begin{enumerate}
    \item $\mathcal{A}_F$ is constructed, 
    \item $\mathcal{A}_F$ is alignment-consistent, and
    \item The MaxSMT/pMaxSMT solver used for repair/verification is complete.
\end{enumerate}
\end{theorem}

Use of MaxSMT guarantees the minimality of edge repair.
\begin{theorem}[Minimality of edge repair]
\label{th:minimality}
Suppose that our algorithm repairs edge $e: u \rightarrow v$ of $\mathcal{A}_F$ by changing $F \subseteq C_s \union U_s$ ($C_s$ and $U_s$ are defined in Definition~\ref{RepairSynthesis}). 
There does not exist $F'$ s.t. $|F'| < |F|$ and the pre-/post-conditions of $e$ are satisfied by replacing the expressions of $F'$ with the expressions in $C_r \union U_r$. 
\end{theorem}

In the following theorem we state the conditions under which the global minimality of our generated repair is guaranteed.

\begin{theorem}[Relative minimality]
A repaired program generated by our algorithm is minimal under the following assumptions:
\begin{enumerate}
    \item Node alignment made in $\mathcal{A}_F$ is optimal in the sense that there is no alternative alignment that can lead to a smaller repair. Note that edge alignment of $\mathcal{A}_F$ is optimal (see Section~\ref{sec:StructuralAlignment}). 
    \item The variable alignment predicates of $\mathcal{A}_F$ are optimal in the sense that there is no alternative variable alignment that can lead to a smaller repair. As shown in Section~\ref{sec:VarAligning}, we use a heuristics-based approach for the sake of efficiency.
\end{enumerate}
\end{theorem}

Despite these limitations, our experimental results show that Verifix tends to find smaller repairs than Clara.
Note that the existing approaches designed to generate minimal repairs~\cite{sarfgen, clara} also do not consider node/edge alignment in the calculation of the minimality of a repair. Instead, a minimal repair is searched for only after node/edge alignment is made.
In fact, unlike those existing approaches that do not consider alignment at all, we consider edge alignment by enumerating over all possible edge alignments between aligned nodes.

\section{Experimental Setup}
\label{sec:experiment_setup}

Evaluation of a programming assignment feedback tool requires a data-set of incorrect student assignments. For our data-set, we chose a publicly released dataset curated by ITSP~\footnote{\url{https://github.com/jyi/ITSP\#dataset-student-programs}}~\cite{yi2017feasibility} for evaluating feasibility of APR techniques on introductory programming assignments. This benchmark consists of incorrect programming assignment submissions by $400+$ first year undergraduate students crediting a \textit{CS-1: Introduction to C Programming} course at a large public university. 
Note that the datasets used by Clara~\cite{clara}, Sarfgen~\cite{sarfgen}, and CoderAssist~\cite{gulwani-fse16} are not publicly released, while the dataset released by Refactory~\cite{refactory} targets Python programming assignments.

We take students' incorrect attempts from four basic weekly programming labs in ITSP benchmark, where each lab consists of several programming assignments that cover different programming topics. For example, the lab in week 3 (Lab-3 in Table~\ref{tab:repair_fse_17}) consists of four programming assignments which teach students about floating-point expressions, printf, and scanf.
Table~\ref{tab:repair_fse_17} lists the four programming labs partitioned by different programming topics. Students had, on average, a time limit of one hour duration for completing each individual assignment. Advanced labs dealing with topics such as multi-dimensional arrays (matrices) and pointers were excluded from our benchmark due to lack of Verification Condition Generation (VCGen) implementation support of converting code to SMT logical formulae. 

Finally, we use $341$ compilable incorrect students' submissions from $28$ various unique programming assignments as our subject.
In addition to the incorrect student submissions, each programming assignment in the ITSP benchmark contains a single reference implementation and a set of test cases designed by the course instructor. 
Both Verifix and baseline Clara~\cite{clara} have access to the reference implementation and test cases to repair the incorrect student programs.

\begin{table*}[t]
    \centering
    {\footnotesize
    \begin{tabular}{l|l|c|c|cc|cc|cc|}
        
        Lab-ID & Topics &\# Assign- & \# Prog- & \multicolumn{2}{c|}{SM (\%)} & \multicolumn{2}{c|}{Repair (\%)} & \multicolumn{2}{c|}{Time (avg)}\\
        & & ments & rams & Clara & Verifix & Clara & Verifix & Clara & Verifix  \\
        \hline
        Lab-3 & Floating point, printf, scanf & 4 & 63 & 100\% & 100\% & 54.0\% & 92.1\% & \M2.1 & 58  \\
        Lab-4 & Conditionals, Simple Loops & 8 & 117 & 92.3\% & 92.6\% & 71.8\% & 82.9\% & 21.0 & 41.5\\ 
        Lab-5 & Nested Loops,  Procedures & 8 & 82 & 24.4\% & 64.6\% & 22.0\% & 45.1\% & 11.1 & 9.7\\ 
        Lab-6 & Integer Arrays & 8 & 79 & 15.2\% & 31.6\% & 12.7\% & 21.5\% & 17.9 & 9.4 \\ 
        \hline
        Total & - & 28 & 341 & 59.5\% & 71.2\% & 42.8\% & 58.4\% & 15.2 & 39\\

        \hline
 
    \end{tabular}
        }
    \caption{Repair accuracy (Repair) and Structural Match (SM) rate of our tool Verifix and Clara~\cite{clara}. Time column represents the average runtime for all successfully repaired programs. Number of assignments in each lab is shown as \textit{\#Assignments}. Number of incorrect student submissions in each lab is shown as \textit{\#Programs}.}
    \label{tab:repair_fse_17}
\end{table*}


\begin{table}[t]
    \centering
    \small
    \begin{tabular}{|lr|lr|}
    \hline
    \multicolumn{2}{|c|}{Clara} & \multicolumn{2}{c|}{Verifix}\\
        Reason & Perc (\%) & Reason & Perc (\%) \\
        \hline         
        Structural Mismatch (SM) & 71\% & Structural Mismatch (SM) & 62\%\\
        Timeout & 27\% & Timeout & 35\%\\
        Unsupported feature & 2\% & SMT issue & 2\%\\
        & & Unsupported feature  & 1\%\\

        \hline
    \end{tabular}
    \caption{Reasons for Clara and Verifix repair failure in Table~\ref{tab:repair_fse_17}}
    \label{tab:repair_failure}
    \vspace{-2em}
\end{table}

\paragraph*{Baseline comparison}
We compare our tool Verifix's performance against the publicly released state-of-art repair tool Clara~\footnote{\url{https://github.com/iradicek/clara}}~\cite{clara} on the common dataset of 341 incorrect student assignments.
A timeout of 5 minutes per incorrect student program was set for both Verifix and Clara to generate the repair. 
We do not directly compare our results against CoderAssist~\cite{gulwani-fse16} tool since it requires additional manual effort in the form of instructor validated submissions, while Refactory~\cite{refactory} implementation targets Python programming assignments. The SarfGen~\cite{sarfgen} tool has not been publicly released and hence cannot be directly compared against. We instead address these comparisons in our related work Section~\ref{sec:related}. 
We will publicly release both our Verifix tool (open-source) and subjects to aid further research.

In principle, Verifix is applicable to any programming language with Verification Condition Generation (VCGen) support; our implementation targets \textit{C} programming assignments.
Our experiments were carried out on a machine with Intel\textsuperscript{\tiny\textregistered} Xeon\textsuperscript{\tiny\textregistered}  E5-2660 v4 @ 2.00 GHz processor and 64 GB of RAM. 

\section{Evaluation}
\label{sec:evaluation}

We address the following research questions in our work.
\begin{enumerate}
    \item RQ1: What is the repair accuracy and reasons for failure of Verifix on the diverse programming assignments dataset?
    \item RQ2: Does Verifix generate small sized meaningful repair?
    \item RQ3: What is the effect of test-suite quality on repair (overfitting)
\end{enumerate}

\subsection{RQ1: Repair accuracy and failure analysis} 
\label{sec:rq1}

Table~\ref{tab:repair_fse_17} compares the repair accuracy of our tool Verifix against the state-of-art tool Clara~\cite{clara} on our dataset of student submissions.
Given a single reference implementation per assignment, Verifix achieves an overall repair accuracy of 58.4\% on the $348$ incorrect programs across $28$ unique assignments. 
In comparison, the baseline tool Clara achieves a lower overall accuracy of 42.5\% on the same assignments, a difference of more than 15\%. 
Note that \emph{the repairs generated by Verifix are verifiably equivalent to the reference implementation}, in addition to passing all the instructor provided test cases. This means that in close to 60\% of the student assignments, we can use Verifix to generate a verifiably correct feedback with confidence, failing which traditional manual feedback generation by human tutors can be used. 

This substantial improvement in repair accuracy of Verifix over Clara is primarily due to the simpler \textit{Structural Match} (SM) requirement of Verifix than that of Clara. 
Verifix requires the nodes in Control-Flow Automata (CFA) structures of both reference and student program to match, in order to generate a repair (refer Section~\ref{sec:StructuralAlignment}). While Clara imposes an additional requirement for the edges of the reference and student CFA to match. That is, in addition to the function and loop structure matching, Clara additionally imposes return/break/continue control-flow matching  between the student and reference program.


This difference in repair accuracy due to structural match failure can be observed from our Table~\ref{tab:repair_fse_17}. 
The reference and student programs for the initial Lab-3 assignments (covering floating-point, printf and scanf topic) are simple; single-procedural without loops. Hence the control flow structure of all the $63$ incorrect student programs matches with the CFG of instructor designed reference program. 
As a result, Clara's CFG based SM algorithm performs similar to Verifix's CFA based SM algorithm, with both achieving 100\% structural match rate.

For comparison, consider the \textit{Lab-5} assignment, involving nested loop and multi-procedures. The difference between Verifix's Structural Match (SM) rate of 64.6\% for Lab-5 and Clara's Structural Match (SM) rate of 24.4\% is \textasciitilde50\%. Due to Verifix's structural matching involving CFA based abstraction, Verifix is able to successfully align 64.6\% of incorrect student programs with reference program, while Clara's stricter CFG based structural matching can align only 24.4\% of Lab-5 incorrect student programs. This allows Verifix to achieve more than twice the repair accuracy of 45.1\% for Lab-5 as compared to Clara's accuracy of 22.0\%.
%
%


 
%

Verifix successfully generates a repair within 39 seconds on average. 
A majority of this time is spent in the invocation of Z3 SMT and pMaxSMT solver for verification and repair purposes, respectively.
In comparison, the average running time of Clara is 15.2 seconds. 
To generate a verifiably correct repair, Verifix uses more time than Clara, suggesting that there is a trade-off to make between correctness guarantee and repair time.
We note that according to an earlier user study~\cite{yi2017feasibility}, students spend 100s on average to resolve semantic errors, and the average running time of Verifix is significantly less than 100s, indicating its possible use in live feedback.

Table~\ref{tab:repair_failure} lists the reasons for repair generation failure by Verifix, along with the weightage of each reason.
\textit{Structural Mismatch} (SM) is the primary reason for repair failure, accounting for  $62\%$ of all the failure cases. A structural mismatch occurs when the loop nesting or number of function definitions do not match between the incorrect program and the single reference solution. 
Since the current version of Verifix does not have the capability to insert/delete loops or functions, we are unable to verify and repair such incorrect student programs. 

The second biggest failure reason is due to \textit{Timeout}, when Verifix exhausts our 5 minutes runtime limit in verifying/repairing the incorrect student submission. This case can occur when the repair space involving reference/incorrect expressions is large. Note that while our algorithm is sound and complete, the z3 solver could struggle with finding a SAT or UNSAT result within our 5 minutes time limit.

In \textasciitilde2\% of the failure cases, Verifix is unable to generate a repair due to SMT issues. These SMT issues mainly occur due to the incomplete theories of integers and reals for non-linear expressions in SMT solver. Leading to an UNKNOWN query results by the SMT solver.
The final issue accounting for \textasciitilde1\% of all failure cases occurs due to lack of support for few programming feature that occur in the dataset. For example, we currently do not support the use of \textit{GOTO} control-flow transitions since they are considered bad programming practice, especially in a pedagogy setting.
From Table~\ref{tab:repair_failure} we note that the failure reasons of Clara follow a similar pattern as Verifix. With Structural Mismatch (SM) and timeout accounting for majority of the failures. Unsupported programs in the dataset accounts for a distant third, at 2\% of all failure cases.

\subsection{RQ2: Minimal repair} 
\label{sec:rq2}

\begin{figure}[t]
    \includegraphics[width=0.5\columnwidth, keepaspectratio]{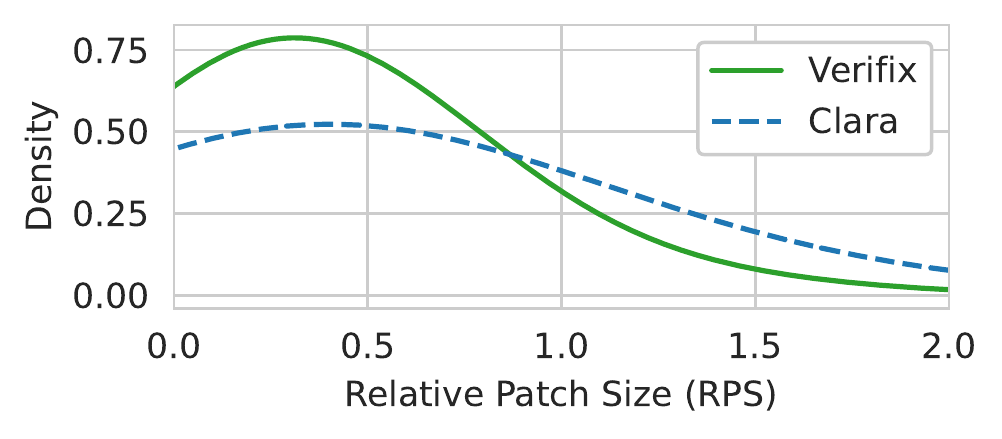}
    \caption{Kernel Density Estimate (KDE) plot of Relative Patch Size (RPS) by Verifix and Clara on 104 common successful repairs.}
    \label{fig:rps}
\end{figure}

\begin{figure}[t] 
{\footnotesize
    \begin{lstlisting}[escapechar=@]
void main(){
  int n1, n2, i;
  scanf("%d %d", &n1, &n2);
  @\hl{if(n2 <= 2)}@        // Repair #1: Delete spurious print
    @\hl{printf("\%d ", n2);}@ //            Verifix @\cmark@, Clara @\xmark@
  for(i=n1; i<=n2; i++){
    if(check_prime(i)@\hl{==0}@) // Repair #2: Delete @\hl{==0}@
      printf("%d ", i);   //   Verifix @\cmark@, Clara @\cmark@ 
  }
}   
    \end{lstlisting}}
  %
    \caption{Example from a Lab-5 \textit{Prime Number} assignment. The main function contains two errors, both of which are fixed by Verifix, while Clara's repair overfits given test-suite by ignoring first error.}
    \label{fig:overfit}
  \end{figure}
In a pedagogy setting, large changes to the incorrect solution can confuse the student.
Hence, apart from correctness measures, we also compare the patch (or repair) size of Verifix and our baseline state-of-art tool Clara~\cite{clara}. 
Since the size of the student programs vary significantly, we normalize patch size with the size of original incorrect program to obtain Relative Patch Size (RPS), given by:
$RPS = {Dist(AST_s, AST_f)} / {Size(AST_s)}$.
Where, $AST_s$ and $AST_f$ represents the Abstract Syntax Tree (AST) of incorrect student program and fixed/repaired program generated by tool, the $Dist$ function computes a \textit{tree-edit-distance} between these ASTs, and the $Size$ function computes the \#nodes in the AST.

In our benchmark of $341$ incorrect programs, Verifix can successfully repair $199$ student programs, Clara can successfully repair $146$ programs, while Verifix and Clara both can successfully repair $104$ common programs.
Out of these $104$ commonly repaired programs, Verifix generates a patch with smaller RPS in $56$ of the cases, Clara generates a patch with smaller RPS in $33$ of the cases, and both tools generate a patch of the exact same relative patch size in $15$ cases.
Note that a smaller repair does not necessarily imply better quality repair since these repairs can overfit the test cases, as we demonstrate in our RQ3 through an example (Section~\ref{sec:rq3}).

Figure~\ref{fig:rps} plots the Kernel Density Estimate (KDE) of Relative Patch Size (RPS) for these $104$ common programs that both Verifix and Clara can successfully repair, in order to visualize the RPS distribution for these large number of data points.
KDE is an estimated Probability Density Function (PDF) of a random variable, often used as a continuous smooth curve replacement for a discrete histogram.
From the Figure~\ref{fig:rps} plot we observe that the density of patch-sizes (y-axis) produced by Verifix is greater than that of Clara when $RPS < 0.8$ (x-axis). 
On the other hand, the density of patch-sizes generated by Clara is greater than that of Verifix when $RPS \geq 0.8$.
That is, a large proportion of repairs generated by Verifix have a small relative patch-size, since the density concentration of repairs is towards lower $RPS$ (x-axis). 
In comparison,  a significantly larger proportion of Clara's repairs have $RPS \geq 0.8$, as compared to Verifix.


\subsection{RQ3: Overfitting} 
\label{sec:rq3}

\begin{figure}[t]
    \includegraphics[width=0.6\columnwidth, keepaspectratio]{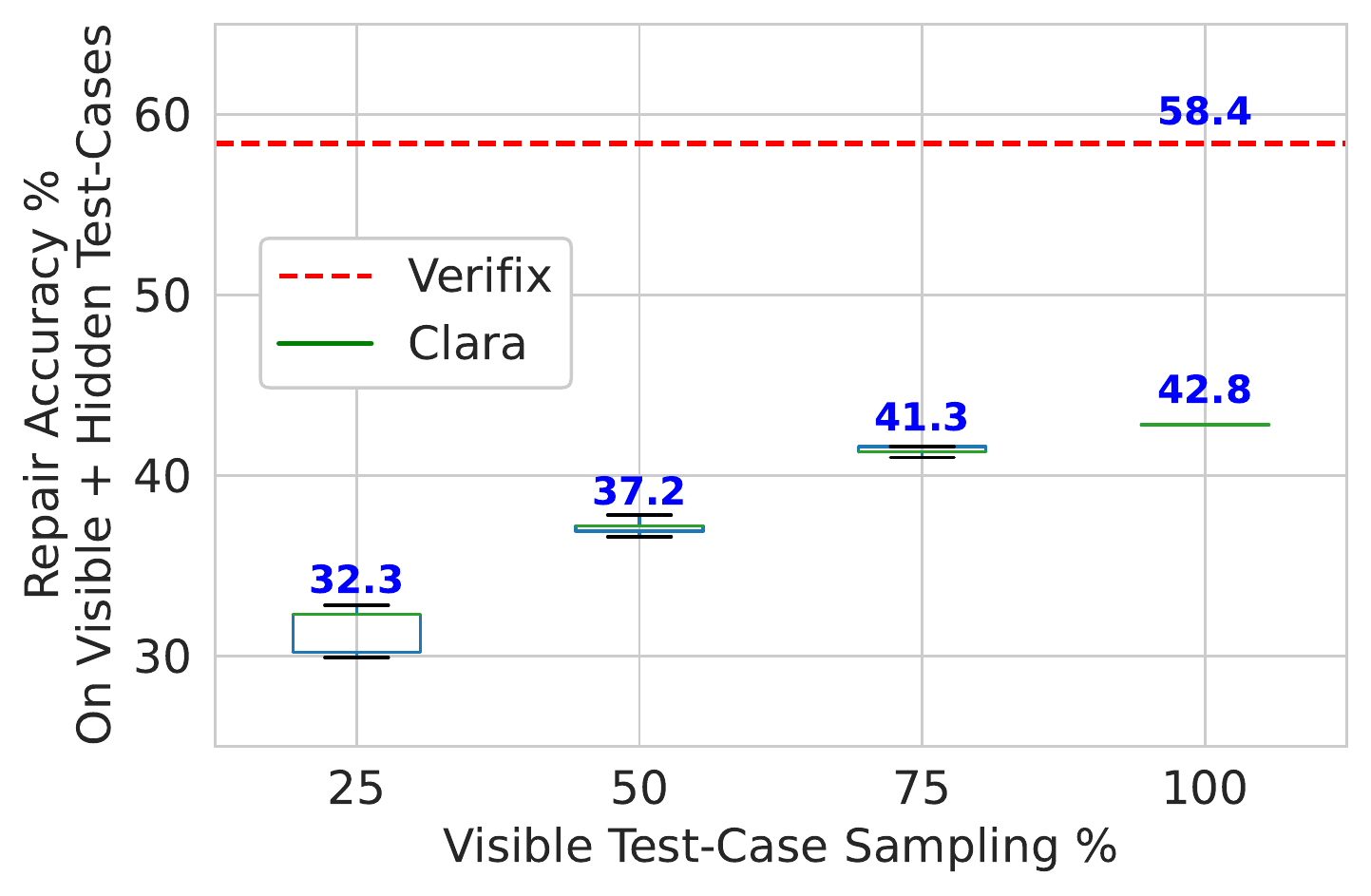}
    \caption{Repair accuracy of Clara and Verifix on various test case samplings.}
    \label{fig:overfit_acc}
\end{figure}


Majority of the programming assignment repair tools~\cite{yi2017feasibility, clara, sarfgen, refactory} generate repairs that satisfy a given test suite (incomplete specification).
Unlike these existing test based approaches, Verifix uses SMT based verification to find a minimal repaired program that behaves similar to a given reference implementation on all possible inputs (complete specification).
Figure~\ref{fig:overfit} demonstrates an example from a Lab-5 \textit{Prime Number} assignment, where Clara's~\cite{clara} repair overfits the test cases.
With the help of a reference implementation, Verifix is able to detect a new counter-example where the student program deviates from correct behavior, when input stream is "1 2" ($n1=1$, $n2=2$).
Given this new unseen test case, the repair suggested by Clara results in an incorrect output "2 2 ", while the repair suggested by Verifix results in the correct behaviour producing output "2 ". 

In order to measure the degree of overfitting repairs generated by each tool, we compare the affect of test case quality on repair accuracy.
This is done by running Clara and Verifix on our common benchmark of $341$ incorrect programs under four different settings, where a percentage of test cases were hidden from tool during repair generation. 
For each of the $28$ unique assignments, with $6$ number of instructor designed test cases on average, we randomly sampled $X\%$ as "visible" test cases.
Once the repair was successfully generated by a tool on the limited visible test case sample, we re-evaluated the repaired program on all test cases, including hidden ones.
We carried out this experiment under four different settings, with a random sampling rate of $25\%$, $50\%$, $75\%$, and $100\%$ of the available test cases.
This entire experiment was repeated 5 times, where we randomly sampled test cases each time, and we report on the distribution of repair accuracy achieved by each tool.

Figure~\ref{fig:overfit_acc} displays the result of our overfitting experiment, with the X-axis representing the visible test case sampling \%, and Y-axis representing the repair accuracy \% obtained by APR tool on the entire test-suite (visible and hidden test cases). 
Each box plot displays the distribution of repair accuracy per test case sampling, by showing the minimum, maximum, upper-quartile, lower-quartile and median values. 
The median value of each box-plot is shown as text above the box-plot. 

From the Figure~\ref{fig:overfit_acc} we observe that Verifix's repair accuracy is constant. That is, Verifix's repair does not change based on the percentage of visible test cases provided, since it does not use the available test cases for repair generation or evaluation/verification.
On the other hand, Clara's repair accuracy varies from a median value of $42.8\%$ (when all test cases are made visible) to a median value of $32.3\%$ (when only $25\%$ of test cases are made available to Clara).
In other words, Clara overfits on $42.8-32.3=10.5\%$ of our benchmark of $341$ incorrect programs, when $25\%$ of test cases are randomly chosen.
Similarly, overfitting of $42.8-41.3=1.5\%$ is observed when visible test case sampling rate is $75\%$, or 5 visible test cases ($\ceil{75\% \times 6}=5$) on average.
In other words, when even a single test case on average is hidden from Clara, its generated repair can overfit the test cases.

Moreover, the choice of test case sampling has a large effect on Clara's accuracy, as evident from the variation in box-plot distribution.
In the case of $25\%$ visible test case sampling, Clara's repair accuracy ranges from a minimum value of $29.9$ to maximum of $32.8$; depending on which two test cases ($\ceil{25\% \times 6}=2$) were made available.

Hence, APR tools such as Clara~\cite{clara} which rely on availability of good quality test cases for their repair generation and evaluation can suffer from overfitting. Even when the instructor misses out on a single important test case coverage during assignment design.
Thereby generating incomplete feedback to students struggling with their incorrect programs.
Verifix on the other hand does not suffer from overfitting limitation, due to its sole reliance on reference implementation for repair generation and evaluation/verification.

\section{Threats to Validity}
\label{sec:threats}

Our aligned automata setup phase consists of a syntactic procedure to obtain a unique edge and variable alignment between the reference and student automata.
Producing an incorrect alignment does not affect our soundness or relative completeness guarantees, but can instead increase our patch-size.
This however occurs rarely in practice, as demonstrated by our RQ2~(Section \ref{sec:rq2}).


The arithmetic theory of SMT solvers is incomplete for non-linear expressions, which can affect our relative completeness. However, this issue affects $<2\%$ of our dataset of incorrect student programs in practice, as demonstrated by our results in Table~\ref{tab:repair_failure}.

Evaluating repair tools using multiple correct student submissions, instead of restricting to a single instructor reference solution, could help improve the repair accuracy.
We mitigate this risk by noting that such an evaluation has been undertaken in the literature earlier~\cite{sarfgen,refactory}, and would benefit both Verifix and our baseline tool Clara in terms of additional Structural Match (SM) rate. Furthermore, we cannot always assume the availability of a large number of reference solutions, in general.

\section{Related Work}
\label{sec:related}


\subsection{General Purpose Program Repair}

Automated Program Repair (APR) \cite{LPR19,Monperrus18} is an enabling technology which allows for the automated fixing of observable program errors thereby relieving the burden of the programmers. 
General purpose APR techniques such as GenProg~\cite{genprog}, SemFix \cite{semfix}, Prophet \cite{prophet} and Angelix~\cite{angelix}, require an incomplete correctness specification typically in the form of a test-suite. 
These techniques achieve low accuracy on student programs that suffer from multiple mistakes, since they can scale to large programs but not necessarily to large repair search spaces~\cite{yi2017feasibility}. 
As student programs are substantially incorrect, the search space of repairs is typically large. 

ITSP~\cite{yi2017feasibility} reports positive results on deploying general APR tools for grading purpose by expert programmers, and negative result when used by novice programmers for feedback. Their low repair accuracy and reliance on test-cases (overfitting) can be seen as a motivator for our work. 

S3~\cite{le2017s3} synthesizes a program using a generic grammar and user-defined test-cases. Semgraft \cite{Semgraft2018} uses simultaneous symbolic execution on a buggy program and a reference program to find a repair, which makes the two program equivalent for a group of test inputs; this class of test inputs is captured by a user-provided input condition. Our work shifts away from test inputs and instead constructs verification guided repair. Furthermore, for our application domain of pedagogy, we seek to build minimal repairs by retaining as much of the buggy program as possible.

Churchill et al~\cite{churchill2019semantic} use trace based alignment to produce a unique aligned automaton, to demonstrate equivalence between two correct programs.  Verifix constructs aligned automata using a syntax-based alignment, with the goal of repairing a given incorrect program to become equivalent to a given correct program.

\subsection{Repair of Programming Assignments}

Autograder~\cite{singh-pldi13} is one of the early approaches in this domain. In Autograder, the correctness of generated patches are verified only in bounded domains (e.g., the size of a list in the program is bounded to a constant number), and thus the verification result is generally unsound.
Autograder also requires instructors to manually provide an error model that specifies common correction patterns of student mistakes, which is not needed in Verifix.

Clara~\cite{clara} too performs bounded unsound verification. 
Clara checks whether each concrete execution trace of the student program matches that of the reference program, and performs a repair on mismatch. Since a concrete execution trace is obtained from test execution, the correctness of a generated patch cannot be guaranteed.
We have provided a detailed experimental comparison with Clara.
Clara assumes the availability of multiple correct student submissions with matching control-flow to the incorrect submissions, limiting their applicability unlike Verifix. 
Sarfgen~\cite{sarfgen} generates patches based on a lightweight syntax-based approach, and assumes the availability of previous student submissions. Both Clara and SarfGen require strict Control-Flow Graph (CFG) similarity between the student and reference program. In comparison, Verifix requires matching function and loop structure between student and reference program. Unlike Clara and SarfGen, Verifix can recover from differences in return/break/continue edge transitions due to its usage of Control-Flow Automata (CFA) based abstraction.

Refactory~\cite{refactory} handles the CFG differences 
by mutating the CFG of the student program to that of the reference program by using a limited set of semantics-preserving refactoring rules, designed manually.
For example, refactoring a while-loop by replacing it with a for-loop structure.
Note that Verifix, unlike Refactory, keeps the original CFG of the student program as much as possible, as shown in Fig~\ref{fig:example}. Our goal is to produce small feedback of high quality. We cannot experimentally compare with Refactory since its implementation targets Python programming assignments.

CoderAssist~\cite{gulwani-fse16}, to the best of our knowledge, is the only \APRS approach that can generate verified feedback.
CoderAssist clusters submissions based on their solution strategy followed by manual identification (or creation) of correct reference solutions in each cluster.
After the clustering phase, CoderAssist undertakes repair at the contract granularity rather than expression granularity --- that is, while CoderAssist can suggest which pre-/post-condition should be met for a code block, CoderAssist does not have the capacity to suggest a concrete expression-level patch.
CoderAssist repair algorithm and evaluation results focus on dynamic programming assignments. In contrast, Verifix is designed and evaluated as a general-purpose \APRS.

There have been several attempts to use neural networks~\cite{bhatia2018neuro,wang2017dynamic,gupta2017deepfix,pu2016sk_p,chhatbar2020macer,tracer2018} for program repair. 
These approaches typically target syntactic/compilation errors, and the repair rate for semantic/logical errors is low~\cite{pu2016sk_p}.
Such machine learning based techniques do not offer any relative completeness guarantees, and their repair soundness is evaluated against an incomplete specification (such as tests).

There has been prior work on live deployment of \APRS tools for repairing student programs~\cite{yi2017feasibility, ahmed2020characterizing}. 
The work of ITSP~\cite{yi2017feasibility} shows negative results on providing semantic repair feedback for student programs. 
At the same time, the work of Tracer~\cite{ahmed2020characterizing} demonstrates positive results for repair based feedback, albeit on simpler (compilation) errors.
In this work, we present an approach for repairing complex logical errors in student programs.  Our tool Verifix can generate verified feedback for 58.4\% of incorrect student submissions for 28 diverse assignments, collected from an actual CS-1 course offering.
The human acceptability of our verified feedback can be further investigated via future user-studies. 

\section{Conclusion}
\label{sec:conclusion}

In this paper, we have presented an approach and tool Verifix, for providing verified repair as feedback to students undertaking introductory programming assignments. The verified repair is generated via relational analysis of the student program and a reference program. Verifix is able to achieve better repair accuracy than existing approaches on our common benchmark. The repairs produced by Verifix are of better quality than state-of-art techniques like Clara \cite{clara}, since they are often smaller in size, while being verifiably equivalent to the instructor provided reference implementation.

\balance
\normalsize

\bibliography{references}
\end{document}